\documentclass[aps,twocolumn,showpacs,preprintnumbers,floats]{revtex4}\def\@cite#1#2{\textsuperscript{[{#1\if@tempswa , #2\fi}]}}

\usepackage{amsmath}
\usepackage{amsfonts}
\usepackage{amssymb}
\usepackage{color}
\usepackage[colorlinks,citecolor=blue,anchorcolor=red,menucolor=red,linkcolor=red,filecolor=red,urlcolor=blue,frenchlinks=red]{hyperref}
\usepackage{epsfig}
\usepackage{graphicx}
\usepackage{graphicx,booktabs}
\usepackage{indentfirst}
\usepackage{longtable,lscape}
\usepackage{makecell}
\usepackage{morefloats}
\usepackage{multirow}
\usepackage{mathrsfs}
\usepackage{subfigure}
\usepackage{txfonts}

\makeatletter
\renewcommand{\maketag@@@}[1]{\hbox{\m@th\normalsize\normalfont#1}}%
\makeatother

\newcommand{\vlab}{\mbox{\boldmath$\lambda$\unboldmath}}

\newcommand{\vxi}{\mbox{\boldmath$\xi$\unboldmath}}

\begin{document}

\title{Higher mass spectra of the fully-charmed and fully-bottom tetraquarks }

\author{ Feng-Xiao Liu$^{1,5}$,
         Ming-Sheng Liu$^{2,5}$~\footnote {E-mail: liumingsheng0001@126.com},
         Xian-Hui Zhong$^{1,5}$~\footnote {E-mail: zhongxh@hunnu.edu.cn},
         Qiang Zhao$^{3,4,5}$~\footnote {E-mail: zhaoq@ihep.ac.cn}}

\affiliation{ 1) Department of Physics, Hunan Normal University, and Key Laboratory of Low-Dimensional Quantum Structures and Quantum Control of Ministry of Education, Changsha 410081, China}
\affiliation{ 2) College of Science, Tianjin University of Technology, Tianjin 300384, China}
\affiliation{ 3) Institute of High Energy Physics, Chinese Academy of Sciences, Beijing 100049, China}
\affiliation{ 4) University of Chinese Academy of Sciences, Beijing 100049, China}
\affiliation{ 5)  Synergetic Innovation Center for Quantum Effects and Applications (SICQEA), Hunan Normal University, Changsha 410081, China}

%\date{\today}

\begin{abstract}

In this work, we calculate the higher mass spectra for the $2S$- and $1D$-wave fully-charmed and fully-bottom tetraquark states in a
nonrelativistic potential quark model. The $2S$-wave fully-charmed/bottom tetraquark states lie
in the mass range of $\sim (6.9,7.1)$/$(19.7,19.9)$ GeV, apart from the highest $0^{++}$ state
$T_{(cc\bar{c}\bar{c})0^{++}}(7185)$/$T_{(bb\bar{b}\bar{b})0^{++}}(19976)$.
Most of the $2S$-wave states highly overlap with the high-lying $1P$-wave states.
The masses for the $1D$-wave fully-charmed/bottom tetraquark states are predicted to be in the range of
$\sim (6.7,7.2)/(19.5,20.0)$ GeV. The mass range for the $D$-wave tetraquark states cover most of
the mass range of the $P$-wave states and the whole mass range of the $2S$-wave states.
The narrow structure $X(6900)$ recently observed at LHCb in the
di-$J/\psi$ invariant mass spectrum may be caused by the $1P$-, or $2S$-, or
$1D$-wave $T_{(cc\bar{c}\bar{c})}$ states. The vague structure $X(7200)$ may be caused by the highest $2S$-wave state
$T_{(cc\bar{c}\bar{c})0^{++}}(7185)$, two low-lying $3S$-wave states $T_{(cc\bar{c}\bar{c})0^{++}}(7240)$
and $T_{(cc\bar{c}\bar{c})2^{++}}(7248)$, and/or the high-lying $1D$-wave states with masses around 7.2 GeV
and $J^{PC}=0^{++},1^{++},2^{++},3^{++}$, or $4^{++}$. While it is apparent that the potential quark model calculations predict more states than the structures observed in the di-$J/\psi$ invariant mass spectrum, our calculations will help further understanding of the properties of these fully-heavy tetraquark states in their strong and magnetic interactions with open channels based on explicit quark model wave functions.

\end{abstract}


\maketitle

\section{Introduction}

Recently, the LHCb Collaboration reported their results on the observations of full-charmed tetraquark $cc\bar{c}\bar{c}$ ($T_{(cc\bar{c}\bar{c})}$) states~\cite{Aaij:2020fnh}. Using the full Run1 and Run2 LHCb data of 9 fb$^{-1}$, the di-$J/\psi$ invariant mass spectrum was studied at $p_T> 5.2$ GeV/c. A broad structure above threshold ranging from 6.2 to 6.8 GeV [denoted by $X(6200-6800)$] and a narrower structure at about 6.9 GeV [denoted by $X(6900)$] are observed with more than 5 $\sigma$ of significance level. There is also a vague structure around 7.2 GeV [denoted by $X(7200)$] to be confirmed. These clear structures may be evidences for genuine tetraquark $T_{(cc\bar{c}\bar{c})}$ states arising from
the quark-gluon interactions in QCD rather than loosely bound hadronic molecules~\cite{Chao:2020dml,Richard:2020hdw}, since the light
mesons cannot be exchanged between two heavy mesons.

Before the LHCb observations, the $T_{(cc\bar{c}\bar{c})}$ spectrum has been widely studied in the literature
~\cite{Chao:1980dv,Ader:1981db,Iwasaki:1975pv,Zouzou:1986qh,Heller:1985cb,Lloyd:2003yc,
Barnea:2006sd,Wang:2017jtz,Karliner:2016zzc,Berezhnoy:2011xn,Bai:2016int,Anwar:2017toa,Esposito:2018cwh,Chen:2016jxd,
Wu:2016vtq,Hughes:2017xie,Richard:2018yrm,Debastiani:2017msn,Wang:2018poa,Richard:2017vry,Vijande:2009kj,Deng:2020iqw,Ohlsson,
Wang:2019rdo,Bedolla:2019zwg,Chen:2020lgj,Chen:2018cqz,Liu:2019zuc}. With the observations at LHCb,
the fully-heavy tetraquark system has again attracted a lot attention from the community. Theoretical studies of various properties of such systems can be found in the literature, such as calculations of the mass spectrum~\cite{liu:2020eha,Karliner:2020dta,Lu:2020cns,Zhao:2020cfi,
Zhao:2020nwy,Gordillo:2020sgc,Jin:2020jfc,
Faustov:2020qfm,Zhang:2020xtb,Yang:2020wkh,Faustov:2021hjs,Yang:2021hrb,
Mutuk:2021hmi,Li:2021ygk,Yang:2020rih,Zhu:2020xni,Wang:2020dlo,Ke:2021iyh,Wang:2020ols,Wang:2021kfv,Majarshin:2021hex,
Weng:2020jao,Giron:2020wpx,Sonnenschein:2020nwn}, production mechanisms~\cite{Huang:2021vtb,Wang:2020gmd,Goncalves:2021ytq,Feng:2020qee,Feng:2020riv,Ma:2020kwb,Maciula:2020wri}, strong decays~\cite{Becchi:2020uvq,Chen:2020xwe}, and analyses of the measured di-$J/\psi$ invariant mass spectrum~\cite{Liang:2021fzr,Dong:2020hxe,Dong:2020nwy,Wang:2020wrp,Gong:2020bmg,Cao:2020gul,Guo:2020pvt} based on different scenarios.

So far, the theoretical interpretations of the LHCb observations are various. For the broad structure $X(6200-6800)$, many analyses based on the mass location~\cite{liu:2020eha,Karliner:2020dta,Wang:2019rdo,Lu:2020cns,Chen:2016jxd,Lloyd:2003yc,Ader:1981db,Zhao:2020cfi,
Zhao:2020nwy,Gordillo:2020sgc,Karliner:2016zzc,Deng:2020iqw,Jin:2020jfc,
Faustov:2020qfm,Zhang:2020xtb,Yang:2020wkh,Faustov:2021hjs,Yang:2021hrb,
Mutuk:2021hmi,Li:2021ygk,Chen:2018cqz,Liu:2019zuc,Chen:2020lgj,Yang:2020rih} indicate that it could be a
good candidate of the ground $T_{(cc\bar{c}\bar{c})}$ states with $J^{PC}=0^{++}$ and $2^{++}$. In contrast, some other models have predicted relatively lower masses than $6200$ MeV for the $0^{++}$ and $2^{++}$ ground states~\cite{Bedolla:2019zwg,Barnea:2006sd,Berezhnoy:2011xn,Wang:2017jtz,Wang:2018poa,Debastiani:2017msn,
Zhu:2020xni,Wang:2020dlo,Ke:2021iyh,Wang:2020ols}. There, the $X(6200-6800)$ may be assigned to some excited states.

For the narrow structure $X(6900)$, theoretical interpretations are far from convergent. In the present literature, $X(6900)$ is suggested to be a candidate of the $2S$-wave (i.e., the first radial excitation) state by several quark model calculations, which include the nonrelativistic potential models~\cite{Wang:2021kfv,Zhao:2020cfi,Zhao:2020nwy}, relativistic quark model~\cite{Faustov:2021hjs}, extended relativized quark model~\cite{Lu:2020cns}, dynamical diquark model~\cite{Giron:2020wpx}, and the string-junction picture~\cite{Karliner:2020dta}. However, it is also assigned as a candidate of some higher $T_{(cc\bar{c}\bar{c})}$ excitations by other models. Its assignments include as a $3S$-wave state~\cite{Zhu:2020xni,Wang:2020dlo,Mutuk:2021hmi,Ke:2021iyh}, $1D$-wave state~\cite{Faustov:2021hjs,Deng:2020iqw}, $1P$-wave state~\cite{liu:2020eha,Chen:2016jxd,Deng:2020iqw,Wang:2021kfv}, and a $2P$-wave state~\cite{Zhu:2020xni}.

Attention is also paid to the vague structure $X(7200)$ in the di-$J/\psi$ invariant mass spectrum~\cite{Aaij:2020fnh}. In the spectrum studies, it is assigned as higher excitation states, such as the second radial $(3S)$ excitations~\cite{Zhao:2020nwy,Faustov:2021hjs}, some $2P$ and/or $2D$ excitations~\cite{Giron:2020wpx}.

Apart from the interpretations which treat these signals as genuine tetraquark states, there are also other explanations proposed in the literature. For instance, Refs.~\cite{Guo:2020pvt,Dong:2020nwy,Wang:2020wrp} describe the structures appearing in the di-$J/\psi$ invariant mass spectrum by coupled-channel effects of double-charmonium rescatterings via an effective potential. In Ref.~\cite{Gong:2020bmg}, it is proposed that the Pomeron exchange between two vector charmonium states can provide strong near-threshold couplings and then dynamically generate pole structures above the two vector charmonium thresholds. Other solutions include treating the narrow $X(6900)$ as a light Higgs-like boson~\cite{Zhu:2020snb}, and $X(6900)$ and $X(7200)$ as gluonic tetracharm states~\cite{Wan:2020fsk}, or $\bar{\Xi}_{cc}\Xi_{cc}$
molecules~\cite{Liu:2020tqy}.

In our previous works~\cite{Liu:2019zuc,liu:2020eha}, we have systematically studied the mass spectra of the $1S$ and $1P$-wave $T_{(cc\bar{c}\bar{c})}$  states within a nonrelativistic potential quark model (NRPQM).
Their masses are predicted to
be within the ranges of $\sim (6445,6550)$ MeV and $\sim(6600, 7000)$ MeV, respectively.
Our results show that the broad structure $X(6200-6800)$ is consistent with the $1S$-wave states around 6.5 GeV, which is similar to the conclusions from Refs.~\cite{liu:2020eha,Karliner:2020dta,Wang:2019rdo,Lu:2020cns,Chen:2016jxd,Lloyd:2003yc,Ader:1981db,Zhao:2020cfi,
Zhao:2020nwy,Gordillo:2020sgc,Karliner:2016zzc,Deng:2020iqw,Jin:2020jfc,
Faustov:2020qfm,Zhang:2020xtb,Yang:2020wkh,Faustov:2021hjs,Yang:2021hrb,
Mutuk:2021hmi,Li:2021ygk,Chen:2018cqz,Liu:2019zuc,Chen:2020lgj,Yang:2020rih}. It should be mentioned that some low-lying $1P$-wave
states with a mass around 6.7 GeV may contribute to $X(6200-6800)$ as well.
Furthermore, it is found that the narrow structure $X(6900)$ may correspond to
some $1P$-wave states around $6.9$ GeV with $J^{PC}=0^{-+},1^{-+},2^{-+}$. Such a possibility was also discussed
by the recent potential quark model calculations~\cite{Deng:2020iqw,Wang:2021kfv}
and the previous QCD sum rule predictions~\cite{Chen:2016jxd}. Actually, since quite a lot of states are predicted by the tetraquark picture, we will see later that some other
possibilities, such as $2S$-, and $1D$-wave states, cannot be eliminated.
The vague structure $X(7200)$ lies about 200 MeV above the high-lying $1P$-wave states
according to our potential quark model predictions~\cite{liu:2020eha}. Given that higher excited states should exist in the quark model, the structure of $X(7200)$ may be an indications of some higher excitations. This is also one of our motivations to investigate the higher $T_{(cc\bar{c}\bar{c})}$ mass spectrum in order to understand the nature of $X(7200)$.

In this work, we continue to study the mass spectra for the higher
$2S$-, $3S$-, and $1D$-wave fully-heavy tetraquark states, i.e. $T_{(QQ\bar{Q}\bar{Q})}$ ($Q=c,b$).
The main purposes are: (i) to investigate the possibility of assigning $X(6900)$ as the $2S$- and/or $1D$-wave
states in the mass spectrum within a
consistent framework; (ii) to know whether or not the $X(7200)$ structure can be associated with the
high $2S$-, $3S$-, and/or $1D$-wave $T_{(cc\bar{c}\bar{c})}$ excitations; (iii) to give relatively
complete $T_{(QQ\bar{Q}\bar{Q})}$ spectra up to the second
orbital excitation. It should be mentioned that many studies of the
spectrum of the radial ($S$-wave) excitations have been carried out~\cite{Wang:2021kfv,Zhao:2020cfi,Zhao:2020nwy,
Faustov:2021hjs, Lu:2020cns,Giron:2020wpx}. However,  because of the complexity of dealing with the four-body system, the study of higher excited states progresses rather slowly. So far, the theoretical predictions on some special $1D$-wave states are based on either the diquark-antidiquark picture, or spin-flavor symmetries as a scheme for parametrization~\cite{Faustov:2021hjs,Bedolla:2019zwg,Deng:2020iqw,Giron:2020wpx}.
For the tetraquark orbital excitations, the spin-orbital and tensor couplings will contribute to the Hamiltonian which will lead to configuration mixings in the wave functions. Thus, the calculation of an orbital excitation tetraquark
system turns out to be more challenging than that for a radial excitation.

The NRPQM adopted in the present work is based on the Hamiltonian
from the Cornell model~\cite{Eichten:1978tg},
which contains a linear confinement potential, a Coulomb-like potential, spin-spin
interactions, spin-orbital interactions, and tensor potentials.
With the NRPQM, we have studied both the $1S$- and $1P$-wave fully-heavy and fully-strange tetraquark states~\cite{Liu:2019zuc,liu:2020eha,Liu:2020lpw} with a
Gaussian expansion method~\cite{Hiyama:2003cu,Hiyama:2018ivm}.
In our calculations, we deal with the fully-heavy tetraquark systems without the
diquark-antidiquark approximation. With the same parameter set as that
for our calculations of the $1S$- and $1P$-wave $T_{(QQ\bar{Q}\bar{Q})}$ ($Q=c,b$) states
in Refs.~\cite{Liu:2019zuc,liu:2020eha}, we obtain self-consistent
predictions of the $T_{(QQ\bar{Q}\bar{Q})}$ mass spectrum for 12 $2S/3S$-wave radial excitations,
and 80 $1D$-wave orbital excitations.

The paper is organized as follows: a brief introduction to the methodology for the tetraquark spectrum
is given in Sec.~\ref{spectrum}. In Sec.~\ref{Results},
the numerical results and discussions are presented. A short summary is given in Sec.~\ref{Summary}.

\section{Methodology}\label{spectrum}

\subsection{Hamiltonian}

We adopt a NRPQM to calculate the masses of the tetraquark states.
In this model, the Hamiltonian is given by
\begin{equation}\label{Hamiltonian}
H=(\sum_{i=1}^4 m_i+T_i)-T_G+\sum_{i<j}V_{ij}(r_{ij}),
\end{equation}
where $m_i$ and $T_i$ stand for the constituent quark mass and kinetic energy of the $i$th quark, respectively; $T_G$ stands for the center-of-mass (c.m.) kinetic energy of the tetraquark system; $r_{ij}\equiv|\mathbf{r}_i-\mathbf{r}_j|$ is the distance between the $i$th and  $j$th quarks; and $V_{ij}(r_{ij})$ stands for the effective potential between them.
In this work the $V_{ij}(r_{ij})$ adopts a widely used form
~\cite{Liu:2020lpw,Li:2020xzs,Liu:2019vtx,Liu:2019wdr,Eichten:1978tg,Godfrey:1985xj,Swanson:2005,Godfrey:2015dia,Godfrey:2004ya,Lakhina:2006fy,Lu:2016bbk,Li:2010vx,Deng:2016stx,Deng:2016ktl}:
\begin{equation}\label{vij}
V_{ij}(r_{ij})=V_{ij}^{conf}(r_{ij})+V_{ij}^{sd}(r_{ij}),
\end{equation}
where the confinement potential adopts the standard form of the Cornell potential~\cite{Eichten:1978tg}, which includes
the spin-independent linear confinement potential $V_{ij}^{Lin}(r_{ij})\propto r_{ij}$ and Coulomb-like potential $V_{ij}^{Coul}(r_{ij})\propto 1/r_{ij}$:
\begin{equation}\label{vcon}
V_{ij}^{conf}(r_{ij})=-\frac{3}{16}({\vlab}_i\cdot{\vlab}_j)\left( b_{ij}r_{ij}-\frac{4}{3}\frac{\alpha_{ij}}{r_{ij}}+C_0 \right).
\end{equation}
The constant $C_0$ stands for the zero point energy. While the spin-dependent potential $V_{ij}^{sd}(r_{ij})$ is the sum of the spin-spin contact
hyperfine potential $V_{ij}^{SS}$, the spin-orbit potential $V_{ij}^{SS}$, and the tensor term $V_{ij}^{T}$:
\begin{equation}\label{voge OGE}
V^{sd}_{ij}(r_{ij})=V^{SS}_{ij}+V^{T}_{ij}+V^{LS}_{ij} \ ,
\end{equation}
with
\begin{equation}\label{voge ss}
V^{SS}_{ij}=-\frac{\alpha_{ij}}{4}({\vlab}_i\cdot{\vlab}_j)\left\{\frac{\pi}{2}\cdot\frac{\sigma^3_{ij}e^{-\sigma^2_{ij}r_{ij}^2}}{\pi^{3/2}}\cdot\frac{16}{3m_im_j}(\mathbf{S}_i\cdot\mathbf{S}_j)\right\},
\end{equation}
\begin{equation}\label{voge ls}
\begin{split}
V^{LS}_{ij}=&-\frac{\alpha_{ij}}{16}\frac{{\vlab}_i\cdot{\vlab}_j}{r_{ij}^3} \bigg(\frac{1}{m_i^2}+\frac{1}{m_j^2}+\frac{4}{m_im_j}\bigg)\bigg\{\mathbf{L}_{ij}\cdot(\mathbf{S}_{i}+\mathbf{S}_{j})\bigg\}\\
&-\frac{\alpha_{ij}}{16}\frac{{\vlab}_i\cdot{\vlab}_j}{r_{ij}^3}\bigg(\frac{1}{m_i^2}-\frac{1}{m_j^2}\bigg)\bigg\{\mathbf{L}_{ij}\cdot(\mathbf{S}_{i}-\mathbf{S}_{j})\bigg\},
\end{split}
\end{equation}
\begin{equation}\label{voge ten}
V^{T}_{ij}=-\frac{\alpha_{ij}}{4}({\vlab}_i\cdot{\vlab}_j)\cdot\frac{1}{m_im_jr_{ij}^3}\Bigg\{\frac{3(\mathbf{S}_i\cdot \mathbf{r}_{ij})(\mathbf{S}_j\cdot \mathbf{r}_{ij})}{r_{ij}^2}-\mathbf{S}_i\cdot\mathbf{S}_j\Bigg\}.
\end{equation}
In the above equations, $\mathbf{S}_i$ stands for the spin of the $i$th quark,
and $\mathbf{L}_{ij}$ stands for the relative orbital angular momentum between the $i$th and $j$th quarks.
If the interaction occurs between two quarks or antiquarks, the $\vlab_i\cdot\vlab_j$ operator is defined as
$\vlab_i\cdot\vlab_j\equiv\sum_{a=1}^8\lambda_i^a\lambda_j^a$, while if the interaction occurs between a quark and an antiquark, the $\vlab_i\cdot\vlab_j$ operator is defined as $\vlab_i\cdot\vlab_j\equiv\sum_{a=1}^8-\lambda_i^a\lambda_j^{a*}$, where $\lambda^{a*}$ is the complex conjugate of the Gell-Mann matrix $\lambda^a$. The parameters $b_{ij}$ and $\alpha_{ij}$ denote the strength of the confinement and strong coupling of the one-gluon-exchange potential, respectively. The quark model parameter sets \{$m_c$, ${\alpha_{cc}}$, ${\sigma_{cc}}$, $b_{cc}$\} and \{$m_b$, ${\alpha_{bb}}$, ${\sigma_{bb}}$, $b_{bb}$\} are taken
the same as those in Refs.~\cite{Deng:2016stx,Liu:2019zuc,liu:2020eha}, they are determined by fitting the charmonium and bottomonium spectra. The quark model parameters adopted in this work are collected in the Table~\ref{parameters}.

%With the wave functions of the tetraquark configurations presented in Tables \ref{states 2S}, ~\ref{states 1D1} and ~\ref{states 1D2}, the mass matrix elements of the Hamiltonian can be worked out.

\begin{table}[htp]
\begin{center}
\caption{\label{parameters} Quark model parameters used in this work.}
\begin{tabular}{cccccccccccc}\hline\hline
%~~~~~~&  $m_u$~(GeV)           &~~~~~~~~~~~~~~~~~~~~~~~~~~~~~~~~~~~~~~0.313~~~~~~~~~~~~~\\
~~~~~~&  $m_c/m_b$~(GeV)           &~~~~~~~~~~~~~~~~~~~~~~~~~~~~~~~~~~~~~~1.483/4.852~~~~~~~~~~~~~\\
~~~~~~&  ${\alpha_{cc}}/\alpha_{bb}$       &~~~~~~~~~~~~~~~~~~~~~~~~~~~~~~~~~~~~~~0.5461/0.4311~~~~~~~~~~~~~\\
~~~~~~&  ${\sigma_{cc}}/\sigma_{bb}$~(GeV) &~~~~~~~~~~~~~~~~~~~~~~~~~~~~~~~~~~~~~~1.1384/2.3200~~~~~~~~~~~~~\\
~~~~~~&  $b_{cc}/b_{bb}$ ~(GeV $^2$)     &~~~~~~~~~~~~~~~~~~~~~~~~~~~~~~~~~~~~~~0.1425/0.1425~~~~~~~~~~~~~\\
\hline\hline
\end{tabular}
\end{center}
\end{table}

%%%%%%%%%%%%%%%%%%%%%%%%%%%%%%%%%%%%%%%%
\subsection{Tetraquark configurations}

The wave function for a $qq\bar{q}\bar{q}$ system can be constructed
as a product of the flavor, color, spin, and spatial configurations.

In the color space, there are two color-singlet bases $|6\bar{6}\rangle_c$ and $|\bar{3}3\rangle_c$, their wave functions are given by
\begin{equation}\label{colourf66}
\begin{split}
|6\bar{6}\rangle_c=&\frac{1}{2\sqrt{6}}\bigg[(rb+br)(\bar{b\mathstrut}\bar{r\mathstrut}+\bar{r\mathstrut}\bar{b\mathstrut})+(gr+rg)(\bar{g\mathstrut}\bar{r\mathstrut}+\bar{r\mathstrut}\bar{g\mathstrut})\\
&+(gb+bg)(\bar{b\mathstrut}\bar{g\mathstrut}+\bar{g\mathstrut}\bar{b\mathstrut})\\
&+2(rr)(\bar{r}\bar{r})+2(gg)(\bar{g}\bar{g})+2(bb)(\bar{b}\bar{b})\bigg],
\end{split}
\end{equation}
\begin{equation}\label{colourf33}
\begin{split}
|\bar{3}3\rangle_c=&\frac{1}{2\sqrt{3}}\bigg[(br-rb)(\bar{b\mathstrut}\bar{r\mathstrut}-\bar{r\mathstrut}\bar{b\mathstrut})-(rg-gr)(\bar{g\mathstrut}\bar{r\mathstrut}-\bar{r\mathstrut}\bar{g\mathstrut})\\
&+(bg-gb)(\bar{b\mathstrut}\bar{g\mathstrut}-\bar{g\mathstrut}\bar{b\mathstrut})\bigg].
\end{split}
\end{equation}

In the spin space, there are six spin bases, which are denoted by $\chi_{SS_z}^{S_{12}S_{34}}$.
Where $S_{12}$ stands for the spin quantum number for the diquark $(q_1q_2)$ (or
antidiquark $(\bar{q}_1\bar{q}_2)$), while $S_{34}$ stands for the spin quantum number
for the antidiquark $(\bar{q}_3\bar{q}_4)$ (or diquark $(q_3q_4)$ ). $S$ is the
total spin quantum number of the tetraquark $qq\bar{q}\bar{q}$ system, while $S_z$ stands for the third component of the total spin $\textbf{S}$.
The spin wave functions $\chi_{SS_z}^{S_{12}S_{34}}$ with a determined $S_z$ can be explicitly expressed as follows:
\begin{eqnarray}\label{spin000}
\chi_{00}^{00}&=&\frac{1}{2}(\uparrow\downarrow\uparrow\downarrow-\uparrow\downarrow\downarrow\uparrow-\downarrow\uparrow\uparrow\downarrow+\downarrow\uparrow\downarrow\uparrow),\\
\chi_{00}^{11}&=&\sqrt{\frac{1}{12}}(2\uparrow\uparrow\downarrow\downarrow-\uparrow\downarrow\uparrow\downarrow-\uparrow\downarrow\downarrow\uparrow\nonumber \\
              &&-\downarrow\uparrow\uparrow\downarrow-\downarrow\uparrow\downarrow\uparrow+2\downarrow\downarrow\uparrow\uparrow),\\
\chi_{11}^{01}&=&\sqrt{\frac{1}{2}}(\uparrow\downarrow\uparrow\uparrow-\downarrow\uparrow\uparrow\uparrow),\\
\chi_{11}^{10}&=&\sqrt{\frac{1}{2}}(\uparrow\uparrow\uparrow\downarrow-\uparrow\uparrow\downarrow\uparrow),\\
\chi_{11}^{11}&=&\frac{1}{2}(\uparrow\uparrow\uparrow\downarrow+\uparrow\uparrow\downarrow\uparrow
-\uparrow\downarrow\uparrow\uparrow-\downarrow\uparrow\uparrow\uparrow),\\
\chi_{22}^{11}&=&\uparrow\uparrow\uparrow\uparrow.
\end{eqnarray}

In the spatial space, we define the relative Jacobi coordinates with the single-partial coordinates
$\mathbf{r_i}$ ($i=1,2,3,4$):
\begin{eqnarray}
\vxi_1&\equiv&\mathbf{r_1}-\mathbf{r_2},\\
\vxi_2&\equiv&\mathbf{r_3}-\mathbf{r_4},\\
\vxi_3&\equiv&\frac{m_1\mathbf{r_1}+m_2\mathbf{r_2}}{m_1+m_2}-\frac{m_3\mathbf{r_3}+m_4\mathbf{r_4}}{m_3+m_4},\\
\mathbf{R}&\equiv&\frac{m_1\mathbf{r_1}+m_2\mathbf{r_2}+m_3\mathbf{r_3}+m_4\mathbf{r_4}}{m_1+m_2+m_3+m_4}.
\end{eqnarray}
Note that $\vxi_1$ and $\vxi_2$ stand for the relative Jacobi coordinates between two quarks
$q_1$ and $q_2$ (or antiquarks $\bar{q}_1$ and $\bar{q}_2$), and two antiquarks $\bar{q}_3$ and $\bar{q}_4$ (or quarks $q_3$ and $q_4$), respectively. While $\vxi_3$ stands for the relative Jacobi coordinate between diquark $qq$ and antidiquark $\bar{q}\bar{q}$.

The spatial bases describing the intrinsic motion of the system with principal
quantum number $N$ and orbital angular momentum quantum numbers $LM$, $\psi_{NLM}$, can be expressed as the linear combination $\psi_{\alpha_1}(\vxi_1)\psi_{\alpha_2}(\vxi_2)\psi_{\alpha_3}(\vxi_3)$:
\begin{eqnarray}\label{cf1}
\psi_{NLM}=\sum_{\alpha_1,\alpha_2,\alpha_3}C_{\alpha_1,\alpha_2,\alpha_3}[\psi_{\alpha_1}(\vxi_1)\psi_{\alpha_2}(\vxi_2)\psi_{\alpha_3}(\vxi_3)]_{NLM},
\end{eqnarray}
where $C_{\alpha_1,\alpha_2,\alpha_3}$ are the combination coefficients.
The functions $\psi_{\alpha_i}(\vxi_i)$ ($i=1,2,3$) stand for the relative-motion wave
functions for the relative Jacobi coordinates $\vxi_i$.
In the quantum number set $\alpha_i\equiv \{n_{\xi_i},l_{\xi_i}, m_{\xi_i}\}$,
$n_{\xi_i}$ is the principal quantum number, $l_{\xi_i}$ is the angular momentum,
and $m_{\xi_i}$ is its third component projection.
For example, for the $1D$-wave states six spatial bases with $M=0$ can be explicitly expressed as
\begin{eqnarray}
|\psi_{020}^{\xi_{1}}\rangle=&\psi_{020}\left(\boldsymbol{\xi}_{1}\right)\psi_{000}\left(\boldsymbol{\xi}_{2}\right)\psi_{000}\left(\boldsymbol{\xi}_{3}\right),\\
|\psi_{020}^{\xi_{2}}\rangle=&\psi_{000}\left(\boldsymbol{\xi}_{1}\right)\psi_{020}\left(\boldsymbol{\xi}_{2}\right)\psi_{000}\left(\boldsymbol{\xi}_{3}\right),\\
|\psi_{020}^{\xi_{3}}\rangle=&\psi_{000}\left(\boldsymbol{\xi}_{1}\right)\psi_{000}\left(\boldsymbol{\xi}_{2}\right)\psi_{020}\left(\boldsymbol{\xi}_{3}\right),\\
|\psi_{020}^{\xi_{1}\xi_{2}}\rangle=&+\frac{1}{\sqrt{6}}\psi_{011}\left(\boldsymbol{\xi}_{1}\right)\psi_{01-1}\left(\boldsymbol{\xi}_{2}\right)\psi_{000}\left(\boldsymbol{\xi}_{3}\right)\nonumber\\
                           &+\sqrt{\frac{2}{3}}\psi_{010}\left(\boldsymbol{\xi}_{1}\right)\psi_{010}\left(\boldsymbol{\xi}_{2}\right)\psi_{000}\left(\boldsymbol{\xi}_{3}\right)\nonumber\\
                           &+\frac{1}{\sqrt{6}}\psi_{01-1}\left(\boldsymbol{\xi}_{1}\right)\psi_{011}\left(\boldsymbol{\xi}_{2}\right)\psi_{000}\left(\boldsymbol{\xi}_{3}\right),\\
|\psi_{010}^{\xi_{1}\xi_{2}}\rangle=&+\frac{1}{\sqrt{2}}\psi_{011}\left(\boldsymbol{\xi}_{1}\right)\psi_{01-1}\left(\boldsymbol{\xi}_{2}\right)\psi_{000}\left(\boldsymbol{\xi}_{3}\right)\nonumber\\
                           &-\frac{1}{\sqrt{2}}\psi_{01-1}\left(\boldsymbol{\xi}_{1}\right)\psi_{011}\left(\boldsymbol{\xi}_{2}\right)\psi_{000}\left(\boldsymbol{\xi}_{3}\right),\\
|\psi_{000}^{\xi_{1}\xi_{2}}\rangle=&+\frac{1}{\sqrt{3}}\psi_{011}\left(\boldsymbol{\xi}_{1}\right)\psi_{01-1}\left(\boldsymbol{\xi}_{2}\right)\psi_{000}\left(\boldsymbol{\xi}_{3}\right)\nonumber\\
                           &-\sqrt{\frac{1}{3}}\psi_{010}\left(\boldsymbol{\xi}_{1}\right)\psi_{010}\left(\boldsymbol{\xi}_{2}\right)\psi_{000}\left(\boldsymbol{\xi}_{3}\right)\nonumber\\
                           &+\frac{1}{\sqrt{3}}\psi_{01-1}\left(\boldsymbol{\xi}_{1}\right)\psi_{011}\left(\boldsymbol{\xi}_{2}\right)\psi_{000}\left(\boldsymbol{\xi}_{3}\right).
\end{eqnarray}
For the $2S$-wave states, the three spatial bases are explicitly expressed as
\begin{eqnarray}
|\psi_{100}^{\xi_{1}}\rangle=&\psi_{100}\left(\boldsymbol{\xi}_{1}\right)\psi_{000}\left(\boldsymbol{\xi}_{2}\right)\psi_{000}\left(\boldsymbol{\xi}_{3}\right),\\
|\psi_{100}^{\xi_{2}}\rangle=&\psi_{000}\left(\boldsymbol{\xi}_{1}\right)\psi_{100}\left(\boldsymbol{\xi}_{2}\right)\psi_{000}\left(\boldsymbol{\xi}_{3}\right),\\
|\psi_{100}^{\xi_{3}}\rangle=&\psi_{000}\left(\boldsymbol{\xi}_{1}\right)\psi_{000}\left(\boldsymbol{\xi}_{2}\right)\psi_{100}\left(\boldsymbol{\xi}_{3}\right).
\end{eqnarray}
The superscript $\xi_i$ ($i=1,2,3$) of the bases stands for an excitation (i.e., $\xi_i$-mode excitation) appearing
within the intrinsic wave function $\psi_{\alpha_i}(\vxi_i)$,
while the superscript $\xi_i \xi_j$ stands for an excitation ($\xi_i \xi_j$-mode excitation) appearing within both
the $\psi_{\alpha_i}(\vxi_i)$ and $\psi_{\alpha_j}(\vxi_j)$ at the same time.

Taking into account the Pauli principle and color confinement for the four-quark system $qq\bar{q}\bar{q}$,
one has 12 configurations for the $2S$-wave radial excitations, and 80 configurations for the $1D$-wave orbital excitations.
The higher radially excited configurations corresponding to the same excited modes of
these $2S$-wave configurations are also easily obtained. The spin-parity quantum numbers, notations, and total wave functions
for these $2S/3S$- and $1D$-wave configurations are presented in Tables~\ref{states 2S},~\ref{states 1D1} and~\ref{states 1D2}.

\subsection{Numerical method}

To work out the matrix elements in the coordinate space, we follow the same method
adopted in our previous works~\cite{Liu:2020lpw,liu:2020eha,Liu:2019zuc}.
As we know, the relative-motion wave functions $\psi_{\alpha_i}(\vxi_i)$ can be expressed as
\begin{equation}\label{cf2}
\psi_{\alpha_i}(\vxi_i)=R_{n_{\xi_i}l_{\xi_i}}(\xi_i) Y_{l_{\xi_i}m_{\xi_i}}(\hat{\vxi_i}),
\end{equation}
where $Y_{l_{\xi_i}m_{\xi_i}}(\hat{\vxi_i})$ are the standard spherical harmonic functions.
The unknown radial parts, $R_{n_{\xi_i} l_{\xi_i}}(\vxi_i)$, are expanded with a series of Gaussian basis functions~\cite{Liu:2019vtx,liu:2020eha}:
\begin{equation}\label{spatial function1}
R_{n_{\xi_i} l_{\xi_i}}(\xi)=\sum_{\ell=1}^n\mathcal{C}_{\xi\ell}~\phi_{n_{\xi_i}l_{\xi_i}}(d_{\xi_i\ell},\vxi_i),
\end{equation}
with
\begin{eqnarray}\label{spatial function2}
%\begin{split}
\phi_{n_{\xi_i}l_{\xi_i}}(d_{\xi_i\ell},\vxi_i)&=\left(\frac{1}{d_{\xi_i\ell}}\right)^{\frac{3}{2}}\Bigg[\frac{2^{l_{\xi_i}+2}}
{(2l_{\xi_i}+1)!!\sqrt{\pi}}\Bigg]^{\frac{1}{2}}\left(\frac{\xi_i}{d_{\xi_i\ell}}\right)^{l_{\xi_i}}e^{-\frac{1}{2}\left(\frac{\xi_i}{d_{\xi_i\ell}}\right)^2}.
%\end{split}
\end{eqnarray}
It should be pointed out that if there are no radial excitations, the expansion method with Gaussian basis functions
are just the same as the expansion method with harmonic oscillator wave
functions. The parameter $d_{\xi_i\ell}$ in Eq.~(\ref{spatial function1}) can be related to the harmonic oscillator frequency $\omega_{\xi_i\ell}$ with $1/d^2_{\xi_i\ell}=\mu_{\xi_i}\omega_{\xi_i\ell}$. For a tetraquark state $T_{(qq\bar{q}\bar{q})}$ containing fully-equal mass quarks,
if we ensure that the spatial wave function with Jacobi coordinates
can transform into the single particle coordinates,
the harmonic oscillator frequencies $\omega_{\xi_i\ell}$ ($i=1,2,3$) can be related to the harmonic oscillator stiffness factor $K_{\ell}$ with $\omega_{\xi_1\ell}=\sqrt{2K_\ell/\mu_{\xi_1}}$, $\omega_{\xi_2\ell}=\sqrt{2K_\ell/\mu_{\xi_2}}$, and $\omega_{\xi_3\ell}=\sqrt{4K_\ell/\mu_{\xi_3}}$.
Taking the reduced masses $\mu_{\xi_1}=\mu_{\xi_2}=m_q/2$, $\mu_{\xi_3}=m_q$ for $T_{(qq\bar{q}\bar{q})}$,
one has $\omega_{\xi_1\ell}=\omega_{\xi_2\ell}=\omega_{\xi_3\ell}=\omega_{\ell}$ and $d_{\xi_i\ell}=(m_{q}/\mu_{\xi_i})^{1/2}d_{\ell}$ with $d_{\ell}=(4m_{q}K_\ell)^{-1/4}$.

\begin{figure}
\centering \epsfxsize=9 cm \epsfbox{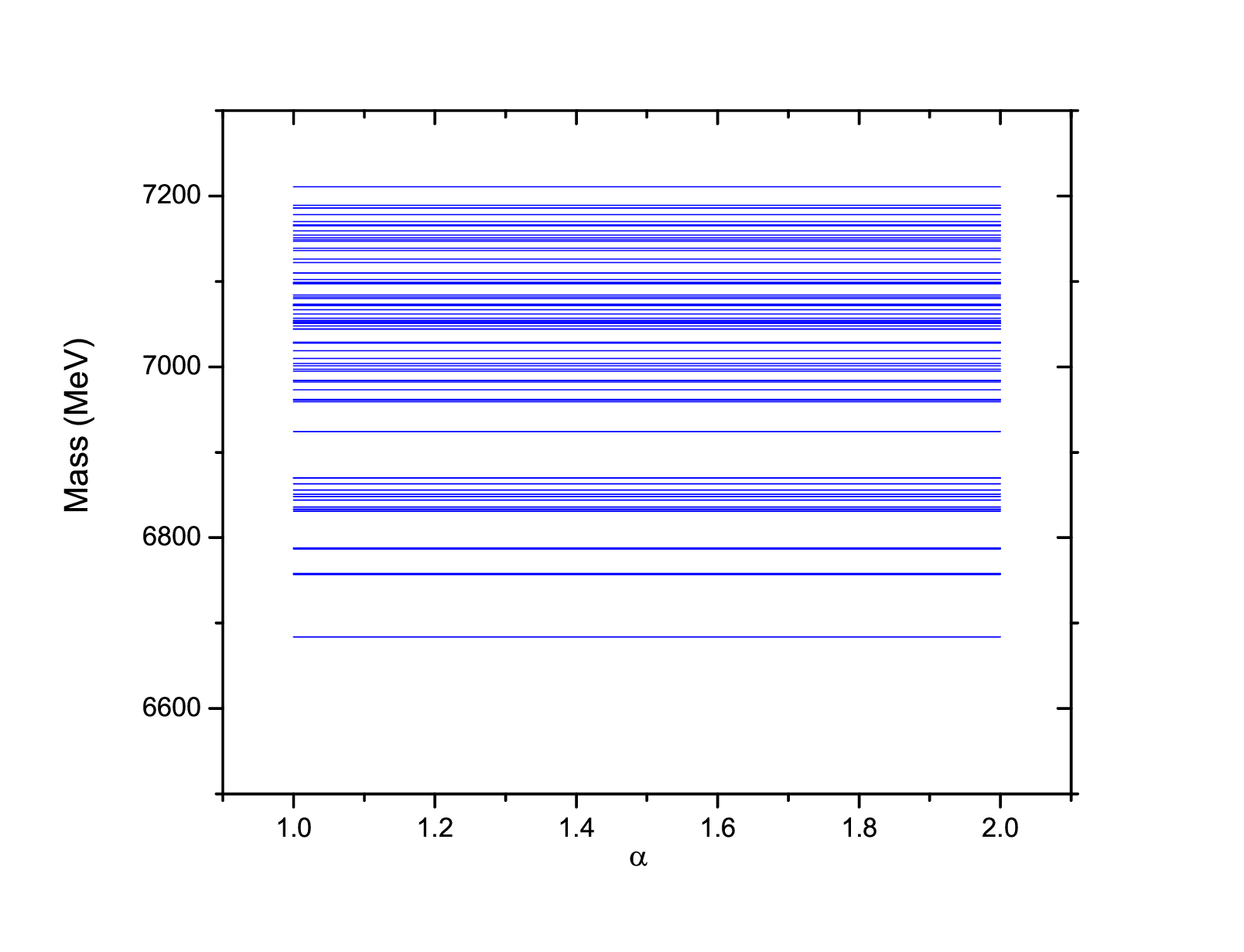} \vspace{-0.9 cm} \caption{Predicted masses of 80 $1D$-wave $T_{(cc\bar{c}\bar{c})}$ configurations as functions of the scaling factor $\alpha$.}\label{figsa}
\end{figure}

Then, the expansion of $\prod_{i=1}^3 R_{n_{\xi_i} l_{\xi_i}}({\xi_i})$ can be simplified as
\begin{equation}
\prod_{i=1}^3 R_{n_{\xi_i} l_{\xi_i}}({\xi_i})\\
=\sum_{\ell}^n \mathcal{C}_{\ell}\prod_{i=1}^3
~\phi_{n_{\xi_i} l_{\xi_i}}(d_{\xi_i\ell},{\xi_i}).
\end{equation}
Following the method of Refs.~\cite{Hiyama:2003cu,Hiyama:2018ivm}, we let the $d_\ell$ parameters form a geometric progression
\begin{equation}\label{geometric progression}
d_\ell=d_1a^{\ell-1}\ \ \ (\ell=1,...,n),
\end{equation}
where $n$ is the number of Gaussian basis functions, and $a$ is the ratio coefficient. There are three parameters $\{d_1,d_n,n\}$ to be determined through the variation method. It is found that with the parameter sets, \{0.068 fm, 2.711 fm, 15\} and \{0.050 fm, 2.016 fm, 15\}, for the $cc\bar{c}\bar{c}$ and $bb\bar{b}\bar{b}$ systems, respectively, we can obtain stable solutions. The numerical results should be independent of the parameter $d_1$. To confirm this point, as done in the literature~\cite{Hiyama:2005cf,Hiyama:2018ukv,Meng:2019fan} we scale the parameter $d_1$ of the basis functions as $d_1\to \alpha d_1$. The mass of a $T_{(QQ\bar{Q}\bar{Q})}$ ($Q=c,b$) state should be stable at a resonance
energy insensitive to the scaling parameter $\alpha$. As an example,
we plot the masses of 80 $1D$-wave $T_{(cc\bar{c}\bar{c})}$ configurations as a
function of the scaling factor $\alpha$ in Fig.~\ref{figsa}.
It is found that the numerical results are nearly independent of the scaling factor $\alpha$.
The stabilization of other states predicted in this work has also been examined by the same method.
With the mass matrix elements ready for every configuration,
the mass of the tetraquark configuration and its spacial wave function can be determined
by solving a generalized eigenvalue problem. The details can be found in our previous works~\cite{Liu:2019vtx,Liu:2019zuc}.
The physical states can be obtained by diagonalizing the mass matrix of different configurations with the same $J^{PC}$ numbers.

Finally, we give some discussions of the numerical method adopted in present work. Usually the trial spatial wave functions
are expanded by a series of Gaussian functions. To keep the completeness of the Gaussian basis set, and to precisely treat an $N$-body system, one can involve several different sets of Jacobi coordinates as those done in Refs.~\cite{Hiyama:2003cu,Hiyama:2018ivm,Meng:2020knc,Hiyama:2005cf,Hiyama:2018ukv,Meng:2019fan},
or adopt a single set of Jacobi coordinates $X=(\xi_1, \xi_2,..., \xi_{N-1})$ with
non diagonal Gaussians $e^{-XAX^T}$ as those done in Refs.~\cite{Varga:1996zz,Varga:1996jr,Varga:1997xga,Brink:1998as}, where $A$ is a symmetric matrix. In our work, we deal with the $cc\bar{c}\bar{c}$ and $bb\bar{b}\bar{b}$ systems which are composed equal mass
constituent quarks and antiquarks. With equal quark masses, a non diagonal term with $\exp\{...+\beta \vxi_i\cdot \vxi_j ...\}$ ($i\neq j$)
implies another one with $\exp\{...-\beta \vxi_i\cdot \vxi_j... \}$, so that the contribution of first order in $\beta$
is eliminated. Thus, the equal mass symmetry can let us select a single set of Jacobi coordinates with
a diagonal Gaussian basis set as an approximation in our study. In this work, the advantage of
adopting a single Jacobi coordinates is that: (i) all the configurations are orthogonal compact multiquark configurations;
(ii) its basis functions of are significantly less than those of several different sets of Jacobi coordinates.
However, with several different sets of Jacobi coordinates, even the continuum states corresponding
to the hadron-hadron scattering solutions come out as discrete states, therefore, one needs adopt a method, such as the real-scaling method, to distinguish the genuine resonances from the discretized scattering states~\cite{Hiyama:2018ukv}.
It should be mentioned that if one deals with a multiquark system composed of mass-different quarks,
a single set of Jacobi coordinates with a diagonal Gaussian basis set may be not a good approximation.

\section{Results and discussions}\label{Results}

Our predicted mass spectra for the $2S/3S$-wave $T_{(cc\bar{c}\bar{c})}$ and $T_{(bb\bar{b}\bar{b})}$ states are
presented in Tables~\ref{mass of cccc2S} and~\ref{mass of bbbb2S}, respectively.
The predictions for the $1D$-wave $T_{(cc\bar{c}\bar{c})}$ states are given in Tables~\ref{mass of cccc1D1}
and~\ref{mass of cccc1D2}, and the predictions for the $1D$-wave $T_{(bb\bar{b}\bar{b})}$ states
are given in Tables \ref{mass of bbbb1D1} and~\ref{mass of bbbb1D2}.
From these tables the components of different configurations for a physical state can be seen.
To see the contributions from each part of the Hamiltonian to the mass of different configurations, we also present our results in Tables~\ref{mass of cccc2Sa}-\ref{mass of bbbb1D2a}.

We find that both the kinetic energy term $\langle T\rangle$ and the linear confinement potential term $\langle V^{Lin}\rangle$ contribute large positive values to the masses, while the Coulomb type potential $\langle V^{Coul}\rangle$ has a large cancellation with these two terms.
The spin-spin interaction term $\langle V^{SS}\rangle$, the tensor potential term $\langle V^{T}\rangle$, and/or the spin-orbit interaction term $\langle V^{LS}\rangle$ have also sizeable contributions to some configurations.
It suggests that a reliable calculation should include both the spin-independent and spin-dependent potentials in the calculations for the tetraquarks. For illustration, our predicted $T_{(cc\bar{c}\bar{c})}$ and $T_{(bb\bar{b}\bar{b})}$ spectra are plotted in Figs.~\ref{figs1} and~\ref{figs2}, respectively.

\subsection{$2S$ and $3S$ states}

In the radially excited ($2S$-, $3S$-wave, etc.) states, apart from the $1S$ ground states with the same quantum numbers, i.e., $J^{PC}=0^{++}, \ 1^{+-}, \ 2^{++}$, some states with additional quantum numbers, i.e., $J^{PC}=0^{+-}, \ 1^{++}, \ 2^{+-}$, can also be accessed.
It should be mentioned that by fully expanding $\prod_{i=1}^3 R_{n_{\xi_i} l_{\xi_i}}({\xi_i})$
with the GEM, one cannot distinguish the $\xi_1$ and $\xi_2$ excitation modes which are
defined for the $2S/3S$ configurations listed in Table~\ref{states 2S}. As a consequence, it is not possible to
numerically work out the masses for the configurations with $J^{PC}=0^{+-}, \ 1^{++}, \ 2^{+-}$. To overcome this problem, following the method adopted in Ref.~\cite{Liu:2020lpw}
we only expand the spatial wave functions containing the radial excitations with series of Gaussian basis functions,
while for those spatial wave functions containing no excitations we adopt the single Gaussian function as an approximation.

There often exist configuration mixings among the states with the
same $J^{PC}$ quantum numbers, which can be seen from the results listed in
Tables~\ref{mass of cccc2S} and ~\ref{mass of bbbb2S}.
For example, there is an obvious mixing between the two $0^{+-}$ states
$2^{1}S_{0^{+-}(6\bar{6})_{c}\left(\xi_{1},\xi_{2}\right)}$ and
$2^{1}S_{0^{+-}(\bar{3}3)_{c}\left(\xi_{1},\xi_{2}\right)}$ in
the $T_{(cc\bar{c}\bar{c})}$ family due to the sizeable off-diagonal
elements contributed by the spin-spin interaction together with the nearly
equal masses for these two $0^{+-}$ configurations.
In contrast, the mixing between
the two $0^{+-}$ $T_{(bb\bar{b}\bar{b})}$ states due to
the spin-spin interaction is strongly suppressed by the heavy bottom quark mass.
The sizeable mixing between the two
$2^{++}$ states $2^{5}S_{2^{++}(\bar{3}3)_{c}\left(\xi_{1},\xi_{2}\right)}$ and
$2^{5}S_{2^{++}(\bar{3}3)_{c}\left(\xi_{3}\right)}$ is mainly caused by the
Coulomb type potential. The Coulomb type potential together
with the linear confinement potential will cause a sizeable
mixing between the two $1^{+-}$ configurations
$2^{3}S_{1^{+-}(\bar{3}3)_{c}\left(\xi_{1},\xi_{2}\right)}$ and
$2^{3}S_{1^{+-}(\bar{3}3)_{c}\left(\xi_{3}\right)}$.

In the configurations with the same $J^{PC}$,
the Coulomb type potentials $\langle V^{Coul}\rangle$ for the $|\bar{3}3\rangle_c$
structure are always more attractive than that for the $|6\bar{6}\rangle_c$ structure.
However, in some cases a $|\bar{3}3\rangle_c$ configuration may have a larger mass than the $|6\bar{6}\rangle_c$ configuration
due to the larger contributions from the linear confinement potential term $\langle V^{Lin}\rangle$
and kinetic energy term $\langle T\rangle$.
For example, for the $T_{(cc\bar{c}\bar{c})}$ sector the $|\bar{6}6\rangle_c$ configuration $2^{1}S_{0^{++}(\bar{6}6)_{c}\left(\xi_{1},\xi_{2}\right)}$	 has a mass of 6954 MeV, which is about 46 MeV smaller than that for
the $|\bar{3}3\rangle_c$ configuration $2^{1}S_{0^{++}(\bar{3}3)_{c}\left(\xi_{1},\xi_{2}\right)}$ of which the mass is 7000 MeV.
It is seen that the contributions $\langle T\rangle=725$ MeV and $\langle V^{Lin}\rangle=883$ MeV for
$2^{1}S_{0^{++}(\bar{6}6)_{c}\left(\xi_{1},\xi_{2}\right)}$ are smaller than
$\langle T\rangle=774$ MeV and $\langle V^{Lin}\rangle=919$ MeV for
$2^{1}S_{0^{++}(\bar{3}3)_{c}\left(\xi_{1},\xi_{2}\right)}$, while $\langle V^{Coul}\rangle=-622$ MeV for $2^{1}S_{0^{++}(\bar{3}3)_{c}\left(\xi_{1},\xi_{2}\right)}$
is more attractive than $\langle V^{Coul}\rangle=-598$ MeV for
$2^{1}S_{0^{++}(\bar{6}6)_{c}\left(\xi_{1},\xi_{2}\right)}$.
It should be mentioned that in the configurations with the same $J^{PC}$, the lowest mass
state is the radial excitation between the diqaurk $(qq)$ and antidiquark $(\bar{q}\bar{q})$  (i.e. the $\xi_3$-mode excitation)
with the $|\bar{3}3\rangle_c$ structure due to the most attractive Coulomb type potential term
$\langle V^{Coul}\rangle$ together with the smallest linear confinement potential term $\langle V^{Lin}\rangle$.

Including configuration mixing effects, the physical masses for the
$2S$ $T_{(cc\bar{c}\bar{c})}$ are predicted to be in the range of
$\sim (6900,7000)$ MeV, except for one $0^{++}$ state
$T_{(cc\bar{c}\bar{c})0^{++}}(7185)$ which has a mass of $M=7185$ MeV.
The masses for the $3S$ $T_{(cc\bar{c}\bar{c})}$ are predicted to be in the range of
$\sim (7200,7400)$ MeV, except for the highest $0^{++}$ state
$T_{(cc\bar{c}\bar{c})0^{++}}(7720)$.
The masses for most of the $2S$ $T_{(bb\bar{b}\bar{b})}$ are predicted to be in the range of
$\sim (19720,19840)$ MeV, except for the highest $0^{++}$ state
$T_{(bb\bar{b}\bar{b})0^{++}}(19976)$. Similarly, the masses for most of the $3S$ $T_{(bb\bar{b}\bar{b})}$ are predicted to be in the range of
$\sim (19980,20130)$ MeV, except for the highest $0^{++}$ state
$T_{(bb\bar{b}\bar{b})0^{++}}(20405)$.
The gap between the lowest $2S$ $T_{(cc\bar{c}\bar{c})}$ state and the highest
$1S$ $T_{(cc\bar{c}\bar{c})}$ state is about $358$ MeV, which is very close to the value $\sim 368$ MeV for
the $T_{(bb\bar{b}\bar{b})}$ sector.
Our predictions for the $2S$ $T_{(cc\bar{c}\bar{c})}$ states are close to those predicted in
Refs.~\cite{Zhao:2020nwy,Lu:2020cns,Wang:2019rdo,Wang:2021kfv}.
Also, our predictions for the $3S$ $T_{(cc\bar{c}\bar{c})}$ states are close to those predicted in
Ref.~\cite{Zhao:2020nwy}. The predicted masses for the $T_{(bb\bar{b}\bar{b})}$ states in our work are
systematically ($\sim 100$ MeV) higher than the results from Refs.~\cite{Zhao:2020nwy,Lu:2020cns}.

Two $2S$ $0^{++}$ states $T_{(cc\bar{c}\bar{c})0^{++}}(6908)$ and $T_{(cc\bar{c}\bar{c})0^{++}}(6957)$ and
one $2^{++}$ state $T_{(cc\bar{c}\bar{c})0^{++}}(6927)$ may be candidates
of the narrow structure $X(6900)$ recently observed at LHCb in the
di-$J/\psi$ invariant mass spectrum~\cite{Aaij:2020fnh}. Their masses are very close to that
of $X(6900)$, and they can decay via a $S$-wave mode $J/\psi J/\psi$.
The highest $0^{++}$ $2S$ state $T_{(cc\bar{c}\bar{c})0^{++}}(7185)$ and one low-lying
$3S$ states $T_{(cc\bar{c}\bar{c})0^{++}}(7240)$ ($3^{3}S_{0^{++}(\bar{3}3)_{c}(\xi_{3})}$)
 may contribute to the vague structure
$X(7200)$ observed at LHCb in the di-$J/\psi$ invariant mass spectrum~\cite{Aaij:2020fnh}.
It should be mentioned that the $2S$ $T_{(cc\bar{c}\bar{c})0^{++}}(7185)$ state is nearly a pure configuration of
$2^{1}S_{0^{++}(6\bar{6})_{c}\left(\xi_{3}\right)}$, with the color structure
$|6\bar{6}\rangle_c$ and the radial excitation
between diquark $(qq)$ and antidiquark $(\bar{q}\bar{q})$. The special color structure
of $T_{(cc\bar{c}\bar{c})0^{++}}(7185)$ leads to a rather large mass gap $\Delta\simeq 167$
MeV from the nearby  $T_{(cc\bar{c}\bar{c})0^{++}}(7018)$. The large mass splitting between
the $2^{1}S_{0^{++}(6\bar{6})_{c}\left(\xi_{3}\right)}$ and the other
$0^{++}$ configurations are also found in the $ss\bar{s}\bar{s}$
and $bb\bar{b}\bar{b}$ families.

Since the predicted masses for the radially excited $T_{(cc\bar{c}\bar{c})}$/$T_{(bb\bar{b}\bar{b})}$
states are far above the low-lying two-charmonium/two-bottomonium mass thresholds,
most of these $2S$ and $3S$ $T_{(cc\bar{c}\bar{c})}$ and $T_{(bb\bar{b}\bar{b})}$
states could be rather broad since they may easily fall apart and decay into two heavy quarkonium channels. Quantitative study of their decays is crucial for better
understanding the properties of these radially excited $T_{(cc\bar{c}\bar{c})}$ and
$T_{(bb\bar{b}\bar{b})}$ states.

\subsection{$1D$ states}

In the 80 $1D$-wave states, apart from the conventional quantum numbers,
i.e., $J^{PC}=0^{++},1^{++},1^{+-},2^{++},3^{++},3^{+-}$ and $4^{++}$, there are exotic
quantum numbers, i.e., $J^{PC}=0^{+-},2^{+-}$ and $4^{+-}$, can be accessed.
From the results listed in Tables~\ref{mass of cccc1D1} and~\ref{mass of cccc1D2},
it shows that there often exist configuration mixings among the states with the
same $J^{PC}$ numbers due to the Coulomb-type potential,
linear-confinement potential, and/or spin-spin interactions.
For example, two $0^{++}$ configurations $1^{3}P_{0^{++}(6\bar{6})_{c}\left(\xi_{1}\xi_{3},\xi_{2}\xi_{3}\right)}$ and $1^{3}P_{0^{++}(\bar{3}3)_{c}\left(\xi_{1}\xi_{3},\xi_{2}\xi_{3}\right)}$ are strongly mixed
with each other by both the Coulomb-type and linear-confinement potentials.
A strong mixing between the two configurations $1^{++}$ $1^{3}S_{1^{++}(6\bar{6})_{c}\left(\xi_{1}\xi_{3},\xi_{2}\xi_{3}\right)}$
and $1^{3}S_{1^{++}(\bar{3}3)_{c}\left(\xi_{1}\xi_{3},\xi_{2}\xi_{3}\right)}$ is mainly
due to the Coulomb-type and spin-spin interactions. It should be mentioned
that the effects of the tensor and the spin-orbit interactions are rather small in the
configuration mixings among the $1D$-wave states due to the suppression of the heavy quark masses.

From Tables~\ref{mass of cccc1D1a}-\ref{mass of bbbb1D2a}, it shows that in
the two $L=0$ (or $L=1$) configurations containing the same $J^{PC}$ quantum numbers
and excitation mode $(\xi_1\xi_3, \xi_2\xi_3)$,
the low mass configuration has a $|6\bar{6}\rangle_c$ structure.
In contrast, in the two $L=2$ configurations containing the same $J^{PC}$ quantum numbers
and excitation mode $(\xi_1\xi_3, \xi_2\xi_3)$,
the low mass configuration has a $|3\bar{3}\rangle_c$ structure.
For example, in the two  $0^{+-}$ $T_{(cc\bar{cc})}$ configurations
$1^{3}P_{0^{+-}(6\bar{6})_{c}\left(\xi_{1}\xi_{3},\xi_{2}\xi_{3}\right)}$
and $1^{3}P_{0^{+-}(3\bar{3})_{c}\left(\xi_{1}\xi_{3},\xi_{2}\xi_{3}\right)}$,
the former has a low mass $6868$ MeV due to the strong attraction from the Coulomb type potential
$\langle V^{Coul}\rangle$ together with the relatively smaller contributions from the linear confinement potential
$\langle V^{Lin}\rangle$. For the two $1^{++}$ $T_{cc\bar{cc}}$ configurations with $L=2$, the mass of $1^{3}D_{1^{++}(6\bar{6})_{c}\left(\xi_{1}\xi_{3},\xi_{2}\xi_{3}\right)}$ ($M=7096$ MeV) turns out to be larger than that of $1^{3}D_{1^{++}(3\bar{3})_{c}\left(\xi_{1}\xi_{3},\xi_{2}\xi_{3}\right)}$
($M=7028$ MeV). It should be mentioned that in the $L=0$ (or $L=1$) configurations,
a $|6\bar{6}\rangle_c$ structure has larger attractions from the Coulomb type potential than
$|3\bar{3}\rangle_c$ although all of the color factors $\langle{\vlab_i}\cdot{\vlab_j}\rangle$
are negative for $|3\bar{3}\rangle_c$. The reason is that for a $|6\bar{6}\rangle_c$ structure the
color factors $\langle{\vlab_1}\cdot{\vlab_3}\rangle=\langle{\vlab_2}\cdot{\vlab_4}\rangle
=\langle{\vlab_1}\cdot{\vlab_4}\rangle=\langle{\vlab_2}\cdot{\vlab_3}\rangle=-10/3$
are a factor of 2.5 larger than $\langle{\vlab_1}\cdot{\vlab_3}\rangle=\langle{\vlab_2}\cdot{\vlab_4}\rangle
=\langle{\vlab_1}\cdot{\vlab_4}\rangle=\langle{\vlab_2}\cdot{\vlab_3}\rangle=-4/3$ for $|3\bar{3}\rangle_c$.

\begin{figure*}
\centering \epsfxsize=17 cm \epsfbox{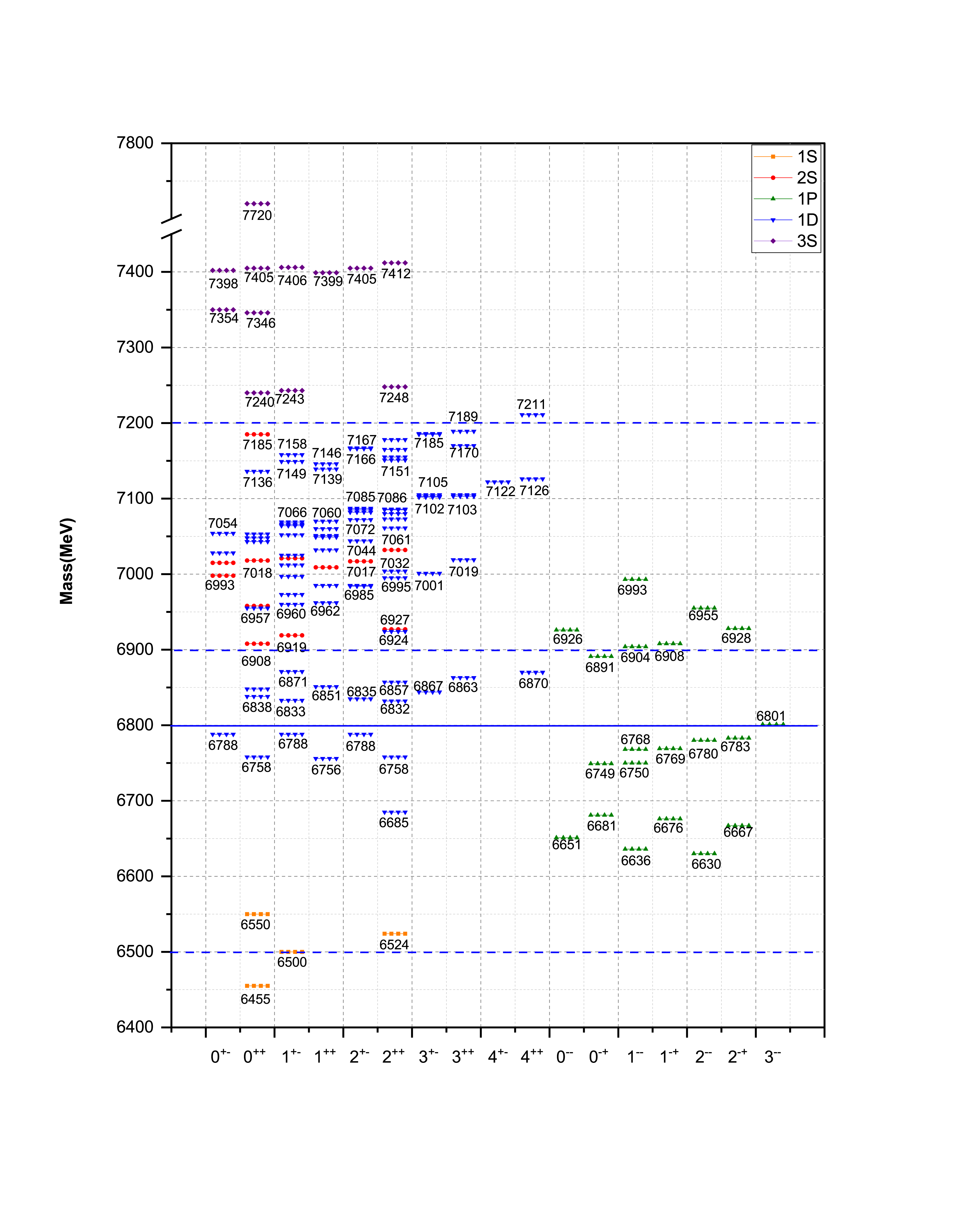} \vspace{-2.5cm} \caption{ $T_{(cc\bar{c}\bar{c})}$ mass spectrum up to the second radial excitations.}\label{figs1}
\end{figure*}

Including configuration mixing effects, the physical masses for the $1D$-wave $T_{(cc\bar{c}\bar{c})}$ states are predicted to be in the range of
$\sim (6700,7200)$ MeV, and the masses for the $1D$-wave $T_{(bb\bar{b}\bar{b})}$ states are predicted to be in the range of
$\sim (19500,20000)$ MeV. The mass range for the $1D$-wave states covers most of
the mass range of the $1P$-wave states and the whole mass range of the $2S$-wave states.
Figure~\ref{figs1} shows that in the mass range $\sim (6700,7000)$ MeV,
many low-lying $1D$-wave $T_{(cc\bar{c}\bar{c})}$ states highly overlap with the $1P$-wave states,
while in the range of $\sim (6900,7050)$ MeV, many $1D$-wave
$T_{(cc\bar{c}\bar{c})}$ states highly overlap with the $2S$-wave states. Such a phenomenon will complicate the experimental analysis if one wants to disentangle their quantum numbers.

For the $T_{(cc\bar{c}\bar{c})}$ sector, the lowest state is the $2^{++}$ state
$T_{(cc\bar{c}\bar{c})2^{++}}(6685)$, which is a mixed state among $1^{1}D_{2^{++}(6\bar{6})_{c}\left(\xi_{1},\xi_{2}\right)}$, $1^{1}D_{2^{++}(\bar{3}3)_{c}\left(\xi_{1}\xi_{2}\right)}$, and
$1^{1}D_{2^{++}(\bar{6}6)_{c}\left(\xi_{3}\right)}$. In contrast, the highest
$T_{(cc\bar{c}\bar{c})}$ state is the $4^{++}$ state $T_{(cc\bar{c}\bar{c})4^{++}}(7211)$.
This state is also a mixed state dominated by the $1^{5}D_{4^{++}(\bar{3}3)_{c}\left(\xi_{3}\right)}$
and $1^{5}D_{4^{++}(\bar{6}6)_{c}\left(\xi_{1}\xi_{2}\right)}$ configurations.
As shown by the numerical results for $T_{(cc\bar{c}\bar{c})2^{++}}(6685)$ and $T_{(cc\bar{c}\bar{c})4^{++}}(7211)$, the configuration mixing effects can strongly shift the masses of the pure configurations to the physical masses.
The mass gap between the lowest mass state $T_{(cc\bar{c}\bar{c})2^{++}}(6685)$ and the highest mass
state $T_{(cc\bar{c}\bar{c})4^{++}}(7211)$ reaches up to a large value of
$\sim 530$ MeV as a combined effect due to the quark interactions. The $\xi_3$ excitation mode is found to mix significantly with the $\xi_1$ and $\xi_2$ excitation modes, and this configuration generally dominates the lowest mass state.

In the $D$-wave $T_{(cc\bar{c}\bar{c})}$ states, three $0^{++}$ states $T_{(cc\bar{c}\bar{c})0^{++}}(6833)$,
$T_{(cc\bar{c}\bar{c})0^{++}}(6848)$ and $T_{(cc\bar{c}\bar{c})0^{++}}(6962)$, two
$1^{++}$ states $T_{(cc\bar{c}\bar{c})1^{++}}(6851)$ and $T_{(cc\bar{c}\bar{c})1^{++}}(6963)$,
three $2^{++}$ states $T_{(cc\bar{c}\bar{c})2^{++}}(6832)$, $T_{(cc\bar{c}\bar{c})2^{++}}(6857)$ and
$T_{(cc\bar{c}\bar{c})2^{++}}(6924)$, one $3^{++}$ state $T_{(cc\bar{c}\bar{c})3^{++}}(6863)$,
and one $4^{++}$ state $T_{(cc\bar{c}\bar{c})4^{++}}(6870)$ may be candidates
of the $X(6900)$ recently observed at LHCb in the
di-$J/\psi$ invariant mass spectrum~\cite{Aaij:2020fnh}. Their masses are close to that
of $X(6900)$, and they can decay into the $J/\psi J/\psi$ channel.
It should be mentioned that the $T_{(cc\bar{c}\bar{c})3^{++}}(6863)$ and
$T_{(cc\bar{c}\bar{c})4^{++}}(6870)$ are higher $J$ states, and decay into the $J/\psi J/\psi$
channel via a $D$-wave mode. Thus, their decays into $J/\psi J/\psi$ could be suppressed by the centrifugal barrier and they may appear to be relatively narrow states.

The vague structure $X(7200)$ observed at LHCb in the di-$J/\psi$ invariant
mass spectrum~\cite{Aaij:2020fnh} may be caused by the high-lying
$D$-wave states with $J^{PC}=0^{++},1^{++},2^{++},3^{++}$, or $4^{++}$.
From Fig.~\ref{figs1} it shows that one $J^{PC}=0^{++}$ state
$T_{(cc\bar{c}\bar{c})0^{++}}(7136)$, two $J^{PC}=1^{++}$ states
$T_{(cc\bar{c}\bar{c})1^{++}}(7141,7146)$, four $J^{PC}=2^{++}$ states
$T_{(cc\bar{c}\bar{c})2^{++}}(7151,7155,7165,7178)$,
two $3^{++}$ states $T_{(cc\bar{c}\bar{c})3^{++}}(7170,7189)$, and one $4^{++}$ state
$T_{(cc\bar{c}\bar{c})4^{++}}(7211)$ just lie in the vicinity of $X(7200)$. Note that some of these states, e.g. $T_{(cc\bar{c}\bar{c})0^{++}}(7136)$, can fall apart and decay into $J/\psi J/\psi$ via the quark rearrangements. This may suggest that the $X(7200)$ may be originated from nontrivial mechanisms.

\section{Summary}\label{Summary}

In this work, we further calculate the higher mass spectra for $2S$-, $3S$-, and $1D$-wave fully charmed and bottom tetraquark states in a nonrelativistic potential quark model, which is a continuation of our previous study of the $1S$- and $1P$-wave states~\cite{Liu:2019zuc,liu:2020eha}.

\begin{figure*}
\centering \epsfxsize=17 cm \epsfbox{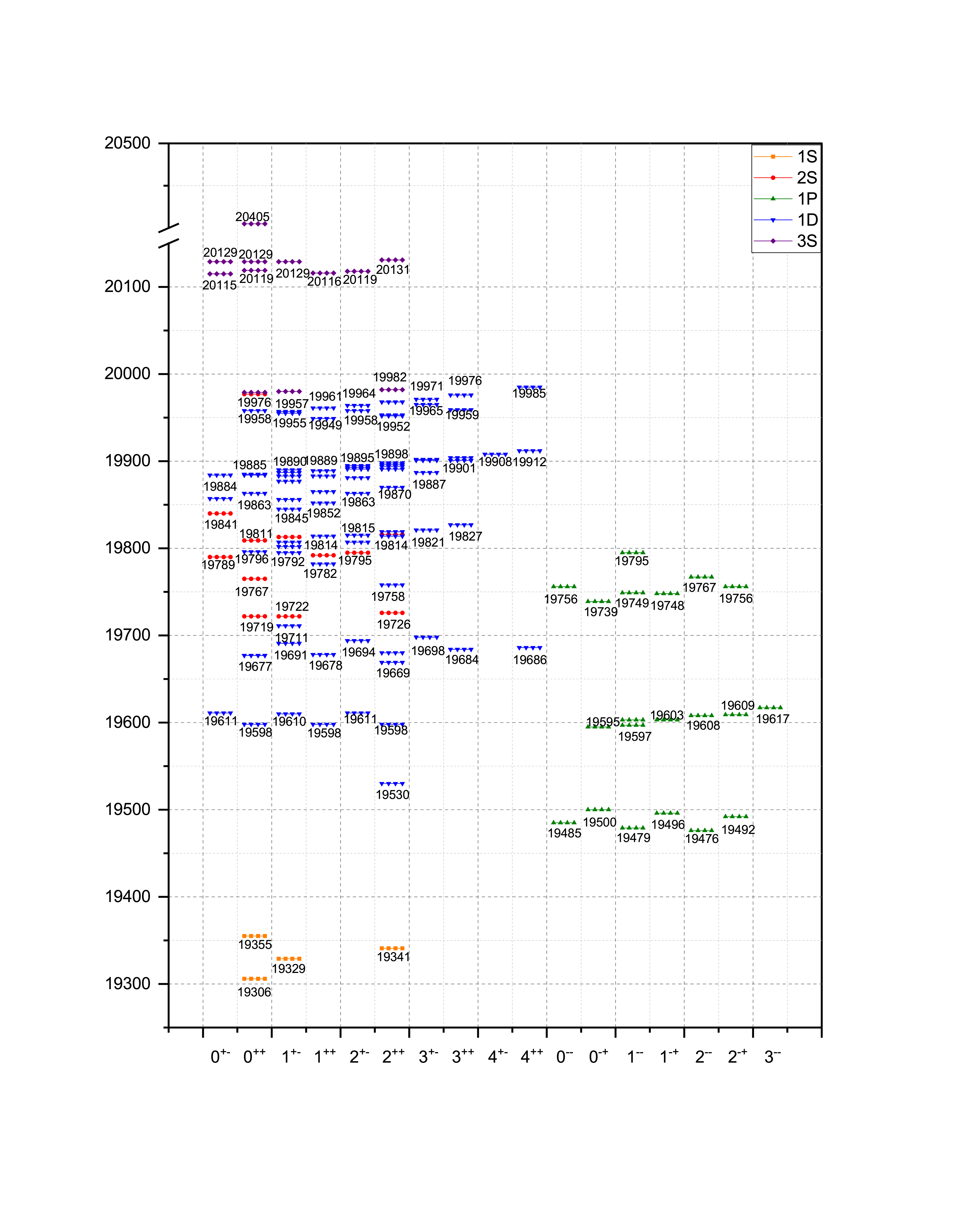} \vspace{-2.5cm} \caption{$T_{(bb\bar{b}\bar{b})}$ mass spectrum up to the second radial excitations.}\label{figs2}
\end{figure*}

Our calculation suggests that within the range of $\sim (6.9,7.2)$ GeV, it scatters the $2S$-wave fully-charmed $T_{(cc\bar{c}\bar{c})}$ states, while the $3S$-wave fully-charmed $T_{(cc\bar{c}\bar{c})}$ states lie in the mass range of $\sim (7.2,7.4)$ GeV with one $0^{++}$ state
$T_{(cc\bar{c}\bar{c})0^{++}}(7720)$ locating at a mass of $M=7720$ MeV. We also find that the masses for the $1D$-wave $T_{(cc\bar{c}\bar{c})}$ states are located in the range of $\sim (6.7,7.2)$ GeV. Notice that this is also the mass region that $1P$-wave states sit~\cite{liu:2020eha}. We actually obtain a busy spectrum for the fully-charmed tetraquark states with the low-excitation quantum numbers. Similar phenomenon occurs for the fully bottomed tetraquark states which can be seen in Fig.~\ref{figs2}.

Our study shows that both the kinetic energy $\langle T\rangle$ and the linear confinement potential $\langle V^{Lin}\rangle$
contribute a large positive value to the tetraquark masses, while the Coulomb type potential $\langle V^{Coul}\rangle$ has a large cancellation with the these two terms. Although some $2S/3S$ and $1D$ configurations have a similar mass, the contributions of $\langle T\rangle$, $\langle V^{Lin}\rangle$ and $\langle V^{Coul}\rangle$ are usually very
different from each other. As a consequence of these interactions, most of the physical states are mixing states with different configurations.

The narrow structure $X(6900)$ may be explained by the $1P$-, or $2S$-, or
$1D$-wave $T_{(cc\bar{c}\bar{c})}$ states. Within the $1P$-wave states, our previous study shows that
$T_{(cc\bar{c}\bar{c})0^{-+}}(6891)$, $T_{(cc\bar{c}\bar{c})1^{-+}}(6908)$, and
$T_{(cc\bar{c}\bar{c})2^{-+}}(6928)$ are possible candidates of $X(6900)$ by looking at the mass locations.
In the sector of the $2S$-wave states, three low-lying states $T_{(cc\bar{c}\bar{c})0^{++}}(6908)$,
$T_{(cc\bar{c}\bar{c})0^{++}}(6957)$, and $T_{(cc\bar{c}\bar{c})2^{++}}(6927)$
may contribute to the narrow structure $X(6900)$. In the $1D$-wave states, three $0^{++}$
states $T_{(cc\bar{c}\bar{c})0^{++}}(6833,6848,6962)$, two
$1^{++}$ states $T_{(cc\bar{c}\bar{c})1^{++}}(6851,6963)$,
three $2^{++}$ states $T_{(cc\bar{c}\bar{c})2^{++}}(6832,6857,6924)$,
one $3^{++}$ state $T_{(cc\bar{c}\bar{c})3^{++}}(6863)$, and one $4^{++}$ state $T_{(cc\bar{c}\bar{c})4^{++}}(6870)$
may be possible candidates for $X(6900)$.

The vague structure $X(7200)$ may be produced by the highest $2S$-wave state
$T_{(cc\bar{c}\bar{c})0^{++}}(7185)$, two low-lying $3S$-wave states $T_{(cc\bar{c}\bar{c})0^{++}}(7240)$
and $T_{(cc\bar{c}\bar{c})2^{++}}(7248)$, or several high-lying $1D$-wave with masses $\sim 7.2$ GeV
and $J^{PC}=0^{++},1^{++},2^{++},3^{++},4^{++}$. It should be mentioned that the $2P$-wave
$T_{(cc\bar{c}\bar{c})}$ states may cover the mass region of $X(7200)$ as well. Unfortunately, in this work we
cannot give predictions of the $2P$-wave spectrum due to its complexity.

To summarize, we have carried out quantitative calculations of the fully-heavy charmed and bottom tetraquark
states in order to gain insights into the four-body system with equal-mass heavy quarks and antiquarks.
We disentangle the important role played by the confinement potential which implies that the fully-heavy
tetraquark states should exist above two heavy quarkonium thresholds. Meanwhile, we find that the potential quark
model will predict an extremely rich spectrum with significant configuration mixings.
This raises questions on the understanding of the experimental observations since apparently only a few states have been seen in
experiment. It should be noted that the rich spectrum predicted in the quark model may be significantly affected by the
strong $S$-wave couplings to the nearby open channels. A combined analysis of the role played by the nearby $S$-wave continuum channels should be the direction for
a better understanding of the multiquark dynamics in the future. Moreover, to uncover the nature of the structures observed in the di-$J/\psi$ invariant mass spectrum, further studies of the decays of these candidates of $T_{(cc\bar{c}\bar{c})}$ states are desired.

%%%%%%%%%%%%%%%%%%%%%%%%%%%%%%%%%%%%%%
\begin{table*}[htp]
\begin{center}
\caption{\label{states 2S} Configurations for the $NS$-wave tetraquark $qq\bar{q}\bar{q}$ system, where $N=n+1$ $(n=1,2,...)$.
$\xi_1,\xi_2,\xi_3$ are the Jacobi coordinates. $(\xi_1,\xi_2)$ stands for a configuration containing both $\xi_1$- and $\xi_2$-mode excitations. }
\centering{}%
\begin{tabular}{clccc}
\hline
\hline
$J^{PC}$ & Configuration & \multicolumn{3}{c}{$\underline{~~~~~\mathrm{Wave}~\mathrm{function}~~~~}$}\tabularnewline
\hline
\multirow{2}{*}{$0^{+-}$} & $N^{1}S_{0^{+-}(6\bar{6})_{c}\left(\xi_{1},\xi_{2}\right)}$ & $\sqrt{\frac{1}{2}}\left(\psi_{n00}^{\xi_{1}}-\psi_{n00}^{\xi_{2}}\right)$ & $\chi_{00}^{00}$ & $|6\bar{6}\rangle^{c}$\tabularnewline
 & $N^{1}S_{0^{+-}(\bar{3}3)_{c}\left(\xi_{1},\xi_{2}\right)}$ & $\sqrt{\frac{1}{2}}\left(\psi_{n00}^{\xi_{1}}-\psi_{n00}^{\xi_{2}}\right)$ & $\chi_{00}^{11}$ & $|\bar{3}3\rangle^{c}$\tabularnewline
\hline
\multirow{4}{*}{$0^{++}$} & $N^{1}S_{0^{++}(6\bar{6})_{c}\left(\xi_{1},\xi_{2}\right)}$ & $\sqrt{\frac{1}{2}}\left(\psi_{n00}^{\xi_{1}}+\psi_{n00}^{\xi_{2}}\right)$ & $\chi_{00}^{00}$ & $|6\bar{6}\rangle^{c}$\tabularnewline
 & $N^{1}S_{0^{++}(\bar{3}3)_{c}\left(\xi_{1},\xi_{2}\right)}$ & $\sqrt{\frac{1}{2}}\left(\psi_{n00}^{\xi_{1}}+\psi_{n00}^{\xi_{2}}\right)$ & $\chi_{00}^{11}$ & $|\bar{3}3\rangle^{c}$\tabularnewline
 & $N^{1}S_{0^{++}(6\bar{6})_{c}\left(\xi_{3}\right)}$ & $\psi_{n00}^{\xi_{3}}$ & $\chi_{00}^{00}$ & $|6\bar{6}\rangle^{c}$\tabularnewline
 & $N^{1}S_{0^{++}(\bar{3}3)_{c}\left(\xi_{3}\right)}$ & $\psi_{n00}^{\xi_{3}}$ & $\chi_{00}^{11}$ & $|\bar{3}3\rangle^{c}$\tabularnewline
\hline
\multirow{2}{*}{$1^{+-}$} & $N^{3}S_{1^{+-}(\bar{3}3)_{c}\left(\xi_{1},\xi_{2}\right)}$ & $\sqrt{\frac{1}{2}}\left(\psi_{n00}^{\xi_{1}}+\psi_{n00}^{\xi_{2}}\right)$ & $\chi_{11}^{11}$ & $|\bar{3}3\rangle^{c}$\tabularnewline
\hline \hline
\end{tabular}
\end{center}
\end{table*}

%%%%%%%%%%%%%%%%%%%%%%%%%%%%%%%%%%%%%%
\begin{table*}[htp]
\begin{center}
\caption{\label{states 1D1} Configurations for the $1D$-wave tetraquark $qq\bar{q}\bar{q}$ system. $\xi_1,\xi_2,\xi_3$ are the Jacobi coordinates.
$(\xi_1,\xi_2)$ stands for a configuration containing both $\xi_1$- and $\xi_2$-mode orbital excitations.
$(\xi_{i}\xi_{j})$ stand for a coupling mode between two different excited modes $\xi_i$ and $\xi_j$.}
\centering{}%
\begin{tabular}{clcc}
\hline\hline
$J^{PC}$ & Configuration & $\underline{~~~~~~~~~~~~~~~\mathrm{Wave}~\mathrm{function}~~~~~~~~~~~~~~~~}$ & \tabularnewline
\hline
\multirow{3}{*}{$0^{+-}$} & $1^{3}P_{0^{+-}(6\bar{6})_{c}\left(\xi_{1}\xi_{3},\xi_{2}\xi_{3}\right)}$ & $\sqrt{\frac{1}{6}}\left(\psi_{011}^{\xi_{1}\xi_{3}}\chi_{1-1}^{10}-\psi_{010}^{\xi_{1}\xi_{3}}\chi_{10}^{10}+\psi_{01-1}^{\xi_{1}\xi_{3}}\chi_{11}^{10}\right)-\sqrt{\frac{1}{6}}\left(\psi_{011}^{\xi_{2}\xi_{3}}\chi_{1-1}^{01}-\psi_{010}^{\xi_{2}\xi_{3}}\chi_{10}^{01}+\psi_{01-1}^{\xi_{2}\xi_{3}}\chi_{11}^{01}\right)$ & $|6\bar{6}\rangle^{c}$\tabularnewline
 & $1^{3}P_{0^{+-}(\bar{3}3)_{c}\left(\xi_{1}\xi_{3},\xi_{2}\xi_{3}\right)}$ & $\sqrt{\frac{1}{6}}\left(\psi_{011}^{\xi_{1}\xi_{3}}\chi_{1-1}^{01}-\psi_{010}^{\xi_{1}\xi_{3}}\chi_{10}^{01}+\psi_{01-1}^{\xi_{1}\xi_{3}}\chi_{11}^{01}\right)-\sqrt{\frac{1}{6}}\left(\psi_{011}^{\xi_{2}\xi_{3}}\chi_{1-1}^{10}-\psi_{010}^{\xi_{2}\xi_{3}}\chi_{10}^{10}+\psi_{01-1}^{\xi_{2}\xi_{3}}\chi_{11}^{10}\right)$ & $|\bar{3}3\rangle^{c}$\tabularnewline
 & $1^{5}D_{0^{+-}(\bar{3}3)_{c}\left(\xi_{1},\xi_{2}\right)}$ & $\begin{array}{cc}
 & \sqrt{\frac{1}{10}}\left(\psi_{022}^{\xi_{1}}\chi_{2-2}^{11}-\psi_{021}^{\xi_{1}}\chi_{2-1}^{11}+\psi_{020}^{\xi_{1}}\chi_{20}^{11}-\psi_{02-1}^{\xi_{1}}\chi_{21}^{11}+\psi_{02-2}^{\xi_{1}}\chi_{22}^{11}\right)\\
- & \sqrt{\frac{1}{10}}\left(\psi_{022}^{\xi_{2}}\chi_{2-2}^{11}-\psi_{021}^{\xi_{2}}\chi_{2-1}^{11}+\psi_{020}^{\xi_{2}}\chi_{20}^{11}-\psi_{02-1}^{\xi_{2}}\chi_{21}^{11}+\psi_{02-2}^{\xi_{2}}\chi_{22}^{11}\right)
\end{array}$ & $|\bar{3}3\rangle^{c}$\tabularnewline
\hline
\multirow{8}{*}{$0^{++}$} & $1^{1}S_{0^{++}(\bar{3}3)_{c}\left(\xi_{1}\xi_{2}\right)}$ & $\psi_{000}^{\xi_{1}\xi_{2}}\chi_{00}^{00}$ & $|\bar{3}3\rangle^{c}$\tabularnewline
 & $1^{1}S_{0^{++}(6\bar{6})_{c}\left(\xi_{1}\xi_{2}\right)}$ & $\psi_{000}^{\xi_{1}\xi_{2}}\chi_{00}^{11}$ & $|6\bar{6}\rangle^{c}$\tabularnewline
\hline\hline
\end{tabular}
\end{center}
\end{table*}

%%%%%%%%%%%%%%%%%%%%%%%%%%%%%%%%%%%%%%%%
\begin{table*}[htp]
\begin{center}
\caption{\label{states 1D2} Configurations for the tetraquark $qq\bar{q}\bar{q}$ system up to the $1D$-wave states (Continued).}
\centering{}%
\begin{tabular}{clcc}
\hline\hline
$J^{PC}$ & Configuration & $\underline{~~~~~~~~~~~~~~~\mathrm{Wave}~\mathrm{function}~~~~~~~~~~~~~~~~}$ & \tabularnewline
\hline
\multirow{11}{*}{$2^{+-}$} & $1^{3}P_{2^{+-}(6\bar{6})_{c}\left(\xi_{1}\xi_{3},\xi_{2}\xi_{3}\right)}$ & \multicolumn{1}{c}{$\sqrt{\frac{1}{2}}\left(\psi_{011}^{\xi_{1}\xi_{3}}\chi_{11}^{10}\right)-\sqrt{\frac{1}{2}}\left(\psi_{011}^{\xi_{2}\xi_{3}}\chi_{11}^{01}\right)$} & $|6\bar{6}\rangle^{c}$\tabularnewline
 & $1^{3}P_{2^{+-}(\bar{3}3)_{c}\left(\xi_{1}\xi_{3},\xi_{2}\xi_{3}\right)}$ & \multicolumn{1}{c}{$\sqrt{\frac{1}{2}}\left(\psi_{011}^{\xi_{1}\xi_{3}}\chi_{11}^{01}\right)-\sqrt{\frac{1}{2}}\left(\psi_{011}^{\xi_{2}\xi_{3}}\chi_{11}^{10}\right)$} & $|\bar{3}3\rangle^{c}$\tabularnewline
 & $1^{5}P_{2^{+-}(6\bar{6})_{c}\left(\xi_{1}\xi_{2}\right)}$ & \multicolumn{1}{c}{$\sqrt{\frac{1}{3}}\psi_{011}^{\xi_{1}\xi_{2}}\chi_{21}^{11}-\sqrt{\frac{2}{3}}\psi_{010}^{\xi_{1}\xi_{2}}\chi_{22}^{11}$} & $|6\bar{6}\rangle^{c}$\tabularnewline
 & $1^{1}D_{2^{+-}(6\bar{6})_{c}\left(\xi_{1},\xi_{2}\right)}$ & \multicolumn{1}{c}{$\sqrt{\frac{1}{2}}\left(\psi_{022}^{\xi_{1}}\chi_{00}^{00}\right)-\sqrt{\frac{1}{2}}\left(\psi_{022}^{\xi_{2}}\chi_{00}^{00}\right)$} & $|6\bar{6}\rangle^{c}$\tabularnewline
 & $1^{1}D_{2^{+-}(\bar{3}3)_{c}\left(\xi_{1},\xi_{2}\right)}$ & \multicolumn{1}{c}{$\sqrt{\frac{1}{2}}\left(\psi_{022}^{\xi_{1}}\chi_{00}^{11}\right)-\sqrt{\frac{1}{2}}\left(\psi_{022}^{\xi_{2}}\chi_{00}^{11}\right)$} & $|\bar{3}3\rangle^{c}$\tabularnewline
 & $1^{3}D_{2^{+-}(\bar{3}3)_{c}\left(\xi_{1},\xi_{2}\right)}$ & $\left(\sqrt{\frac{2}{6}}\psi_{022}^{\xi_{1}}\chi_{10}^{11}-\sqrt{\frac{1}{6}}\psi_{021}^{\xi_{1}}\chi_{11}^{11}\right)+\left(\sqrt{\frac{2}{6}}\psi_{022}^{\xi_{2}}\chi_{10}^{11}-\sqrt{\frac{1}{6}}\psi_{021}^{\xi_{2}}\chi_{11}^{11}\right)$ & $|\bar{3}3\rangle^{c}$\tabularnewline
\hline\hline
\end{tabular}
\end{center}
\end{table*}

%%%%%%%%%%%%%%%%%%%%%%%%%%%%%%%%%%%%%%%%
\begin{table*}[htp]
\begin{center}
\caption{\label{mass of cccc2S} Predicted mass spectrum for the $2S$- and $3S$-wave $T_{(cc\bar{c}\bar{c})}$ states.}
\begin{tabular}{ccccc}
\hline\hline
$J^{PC}$ & Configuration & $\langle H\rangle$ (MeV) & Mass (MeV) & Eigenvector\tabularnewline
\hline
$0^{+-}$ & $\begin{array}{l}
2^{1}S_{0^{+-}(6\bar{6})_{c}\left(\xi_{1},\xi_{2}\right)}\\
2^{1}S_{0^{+-}(\bar{3}3)_{c}\left(\xi_{1},\xi_{2}\right)}
\end{array}$ & $\left(\begin{array}{cc}
7008 & -14\\
-14 & 7005
\end{array}\right)$ & $\left(\begin{array}{c}
6993\\
7020
\end{array}\right)$ & $\left(\begin{array}{cc}
-0.66 & -0.75\\
-0.75 & 0.66
\end{array}\right)$\tabularnewline
$0^{++}$ & $\begin{array}{l}
2^{1}S_{0^{++}(6\bar{6})_{c}\left(\xi_{1},\xi_{2}\right)}\\
2^{1}S_{0^{++}(\bar{3}3)_{c}\left(\xi_{1},\xi_{2}\right)}\\
2^{1}S_{0^{++}(6\bar{6})_{c}\left(\xi_{3}\right)}\\
2^{1}S_{0^{++}(\bar{3}3)_{c}\left(\xi_{3}\right)}
\end{array}$ & $\left(\begin{array}{cccc}
6954 & -19 & 13 & -11\\
-19 & 7000 & -4 & -36\\
13 & -4 & 7183 & -12\\
-11 & -36 & -12 & 6930
\end{array}\right)$ & $\left(\begin{array}{c}
6908\\
6957\\
7018\\
7185
\end{array}\right)$ & $\left(\begin{array}{cccc}
0.35 & 0.40 & 0.03 & 0.84\\
-0.91 & -0.05 & 0.07 & 0.41\\
0.21 & -0.91 & -0.01 & 0.35\\
-0.06 & 0.02 & -1.00 & 0.05
\end{array}\right)$\tabularnewline
$1^{+-}$ & $\begin{array}{l}
2^{3}S_{1^{+-}(\bar{3}3)_{c}\left(\xi_{1},\xi_{2}\right)}\\
2^{3}S_{1^{+-}(\bar{3}3)_{c}\left(\xi_{3}\right)}
\end{array}$ & $\left(\begin{array}{cc}
7006 & -37\\
-37 & 6934
\end{array}\right)$ & $\left(\begin{array}{c}
6919\\
7021
\end{array}\right)$ & $\left(\begin{array}{cc}
-0.39 & -0.92\\
-0.92 & 0.39
\end{array}\right)$\tabularnewline
$1^{++}$ & $\begin{array}{l}
2^{3}S_{1^{++}(\bar{3}3)_{c}\left(\xi_{1},\xi_{2}\right)}\end{array}$ & $\left(\begin{array}{c}
7009\end{array}\right)$ & $7009$ & 1\tabularnewline
$2^{+-}$ & $\begin{array}{l}
2^{5}S_{2^{+-}(\bar{3}3)_{c}\left(\xi_{1},\xi_{2}\right)}\end{array}$ & $\left(\begin{array}{c}
7017\end{array}\right)$ & $7017$ & 1\tabularnewline
\hline\hline
\end{tabular}%
\end{center}
\end{table*}

%%%%%%%%%%%%%%%%%%%%%%%%%%%%%%%%%%%%%%%%
\begin{table*}[htp]
\begin{center}
\caption{\label{mass of bbbb2S} Predicted mass spectrum for the $2S$- and $3S$-wave $T_{(bb\bar{b}\bar{b})}$ states.}
\begin{tabular}{ccccc}
\hline\hline
$J^{PC}$ & Configuration & $\langle H\rangle$ (MeV) & Mass (MeV) & Eigenvector\tabularnewline
\hline
$0^{+-}$ & $\begin{array}{l}
2^{1}S_{0^{+-}(6\bar{6})_{c}\left(\xi_{1},\xi_{2}\right)}\\
2^{1}S_{0^{+-}(\bar{3}3)_{c}\left(\xi_{1},\xi_{2}\right)}
\end{array}$ & $\left(\begin{array}{cc}
19840 & 5\\
5 & 19790
\end{array}\right)$ & $\left(\begin{array}{c}
19789\\
19841
\end{array}\right)$ & $\left(\begin{array}{cc}
0.10 & -1.00\\
-1.00 & -0.10
\end{array}\right)$\tabularnewline
$0^{++}$ & $\begin{array}{l}
2^{1}S_{0^{++}(6\bar{6})_{c}\left(\xi_{1},\xi_{2}\right)}\\
2^{1}S_{0^{++}(\bar{3}3)_{c}\left(\xi_{1},\xi_{2}\right)}\\
2^{1}S_{0^{++}(6\bar{6})_{c}\left(\xi_{3}\right)}\\
2^{1}S_{0^{++}(\bar{3}3)_{c}\left(\xi_{3}\right)}
\end{array}$ & $\left(\begin{array}{cccc}
19770 & -7 & -30 & -4\\
-7 & 19797 & -1 & -31\\
-30 & -1 & 19972 & -4\\
-4 & -31 & -4 & 19733
\end{array}\right)$ & $\left(\begin{array}{c}
19719\\
19767\\
19811\\
19976
\end{array}\right)$ & $\left(\begin{array}{cccc}
0.15 & 0.38 & 0.04 & 0.91\\
-0.97 & -0.04 & -0.14 & 0.19\\
0.10 & -0.92 & 0.02 & 0.37\\
0.14 & 0 & -0.99 & 0.01
\end{array}\right)$\tabularnewline
$1^{+-}$ & $\begin{array}{l}
2^{3}S_{1^{+-}(\bar{3}3)_{c}\left(\xi_{1},\xi_{2}\right)}\\
2^{3}S_{1^{+-}(\bar{3}3)_{c}\left(\xi_{3}\right)}
\end{array}$ & $\left(\begin{array}{cc}
19800 & -32\\
-32 & 19735
\end{array}\right)$ & $\left(\begin{array}{c}
19722\\
19813
\end{array}\right)$ & $\left(\begin{array}{cc}
-0.38 & -0.92\\
-0.92 & 0.38
\end{array}\right)$\tabularnewline
$1^{++}$ & $\begin{array}{l}
2^{3}S_{1^{++}(\bar{3}3)_{c}\left(\xi_{1},\xi_{2}\right)}\end{array}$ & $\left(\begin{array}{c}
19792\end{array}\right)$ & 19792 & 1\tabularnewline
$2^{+-}$ & $\begin{array}{l}
2^{5}S_{2^{+-}(\bar{3}3)_{c}\left(\xi_{1},\xi_{2}\right)}\end{array}$ & $\left(\begin{array}{c}
19795\end{array}\right)$ & 19795 & 1\tabularnewline
\hline\hline
\end{tabular}%
\end{center}
\end{table*}

%%%%%%%%%%%%%%%%%%%%%%%%%%%%%%%%%%%%%%%%
\begin{table*}[htp]
\begin{center}
\caption{\label{mass of cccc1D1} Predicted mass spectrum for the $1D$-wave $T_{(cc\bar{c}\bar{c})}$ states.}
\begin{tabular}{ccccc}
\hline\hline
$J^{PC}$ & Configuration & $\langle H\rangle$ (MeV) & Mass (MeV) & Eigenvector\tabularnewline
\hline
\multirow{2}{*}{$0^{+-}$} & $\begin{array}{l}
1^{3}P_{0^{+-}(6\bar{6})_{c}\left(\xi_{1}\xi_{3},\xi_{2}\xi_{3}\right)}\\
1^{3}P_{0^{+-}(\bar{3}3)_{c}\left(\xi_{1}\xi_{3},\xi_{2}\xi_{3}\right)}
\end{array}$ & $\left(\begin{array}{cc}
6868 & -113\\
-113 & 6948
\end{array}\right)$ & $\left(\begin{array}{c}
6788\\
7028
\end{array}\right)$ & $\left(\begin{array}{cc}
-0.82 & -0.58\\
-0.58 & 0.82
\end{array}\right)$\tabularnewline
 & $1^{5}D_{0^{+-}(\bar{3}3)_{c}\left(\xi_{1},\xi_{2}\right)}$ & $\left(\begin{array}{c}
7054\end{array}\right)$ & 7054 & 1\tabularnewline
\hline
\multirow{5}{*}{$0^{++}$} & $1^{1}S_{0^{++}(\bar{3}3)_{c}\left(\xi_{1}\xi_{2}\right)}$ & $\left(6838\right)$ & 6838 & 1\tabularnewline
 & $1^{1}S_{0^{++}(6\bar{6})_{c}\left(\xi_{1}\xi_{2}\right)}$ & $\left(6957\right)$ & 6957 & 1\tabularnewline
 & $\begin{array}{l}
1^{3}P_{0^{++}(6\bar{6})_{c}\left(\xi_{1}\xi_{3},\xi_{2}\xi_{3}\right)}\\
1^{3}P_{0^{++}(\bar{3}3)_{c}\left(\xi_{1}\xi_{3},\xi_{2}\xi_{3}\right)}
\end{array}$ & $\left(\begin{array}{cc}
6857 & 136\\
136 & 6944
\end{array}\right)$ & $\left(\begin{array}{c}
6758\\
7043
\end{array}\right)$ & $\left(\begin{array}{cc}
-0.81 & 0.59\\
0.59 & 0.81
\end{array}\right)$\tabularnewline
 & $1^{3}P_{0^{++}(6\bar{6})_{c}\left(\xi_{1}\xi_{2}\right)}$ & $\left(7053\right)$ & 7053 & 1\tabularnewline
 & $\begin{array}{l}
1^{5}D_{0^{++}(\bar{3}3)_{c}\left(\xi_{1},\xi_{2}\right)}\\
1^{5}D_{0^{++}(\bar{3}3)_{c}\left(\xi_{3}\right)}\\
1^{5}D_{0^{++}(6\bar{6})_{c}\left(\xi_{1}\xi_{2}\right)}
\end{array}$ & $\left(\begin{array}{ccc}
7051 & 61 & 6\\
61 & 6968 & -128\\
6 & -128 & 7013
\end{array}\right)$ & $\left(\begin{array}{c}
7048\\
6848\\
7136
\end{array}\right)$ & $\left(\begin{array}{ccc}
0.88 & -0.09 & 0.47\\
-0.25 & 0.76 & 0.6\\
0.41 & 0.64 & -0.65
\end{array}\right)$\tabularnewline
\hline\hline
\end{tabular}
\end{center}
\end{table*}

%%%%%%%%%%%%%%%%%%%%%%%%%%%%%%%%%%%%%%%%
\begin{table*}[htp]
\begin{center}
\caption{\label{mass of cccc1D2} Predicted mass spectrum for the $1D$-wave $T_{(cc\bar{c}\bar{c})}$ states (Continued).}
\begin{tabular}{ccccc}
\hline\hline
$J^{PC}$ & Configuration & $\langle H\rangle$ (MeV) & Mass (MeV) & Eigenvector\tabularnewline
\hline
\multirow{7}{*}{$2^{++}$} & $1^{5}S_{2^{++}(6\bar{6})_{c}\left(\xi_{1}\xi_{2}\right)}$ & $\left(7004\right)$ & 7004 & 1\tabularnewline
 & $\begin{array}{l}
1^{3}P_{2^{++}(6\bar{6})_{c}\left(\xi_{1}\xi_{3},\xi_{2}\xi_{3}\right)}\\
1^{3}P_{2^{++}(\bar{3}3)_{c}\left(\xi_{1}\xi_{3},\xi_{2}\xi_{3}\right)}
\end{array}$ & $\left(\begin{array}{cc}
6864 & 145\\
145 & 6955
\end{array}\right)$ & $\left(\begin{array}{c}
6758\\
7061
\end{array}\right)$ & $\left(\begin{array}{cc}
-0.81 & 0.59\\
0.59 & 0.81
\end{array}\right)$\tabularnewline
 & $1^{3}P_{2^{++}(6\bar{6})_{c}\left(\xi_{1}\xi_{2}\right)}$ & $\left(7073\right)$ & 7073 & 1\tabularnewline
 & $\begin{array}{l}
1^{1}D_{2^{++}(6\bar{6})_{c}\left(\xi_{1},\xi_{2}\right)}\\
1^{1}D_{2^{++}(\bar{3}3)_{c}\left(\xi_{1},\xi_{2}\right)}\\
1^{1}D_{2^{++}(6\bar{6})_{c}\left(\xi_{3}\right)}\\
1^{1}D_{2^{++}(\bar{3}3)_{c}\left(\xi_{3}\right)}\\
1^{1}D_{2^{++}(\bar{3}3)_{c}\left(\xi_{1}\xi_{2}\right)}\\
1^{1}D_{2^{++}(6\bar{6})_{c}\left(\xi_{1}\xi_{2}\right)}
\end{array}$ & $\left(\begin{array}{cccccc}
6955 & -19 & 174 & 5 & 25 & 3\\
-19 & 7073 & 5 & 72 & 1 & 28\\
174 & 5 & 6964 & -21 & -147 & -6\\
5 & 72 & -21 & 6986 & -2 & -153\\
25 & 1 & -147 & -2 & 6859 & -20\\
3 & 28 & -6 & -153 & -20 & 7018
\end{array}\right)$ & $\left(\begin{array}{c}
7151\\
7086\\
6685\\
6832\\
6924\\
7178
\end{array}\right)$ & $\left(\begin{array}{cccccc}
-0.45 & -0.17 & -0.55 & -0.45 & 0.21 & 0.46\\
0.03 & -0.9 & -0.08 & 0.02 & 0.09 & -0.42\\
0.46 & 0.06 & -0.62 & -0.13 & -0.6 & -0.12\\
-0.09 & 0.28 & 0.1 & -0.72 & 0.13 & -0.61\\
-0.65 & -0.08 & 0.21 & 0.01 & -0.72 & -0.07\\
-0.38 & 0.28 & -0.5 & 0.51 & 0.23 & -0.46
\end{array}\right)$\tabularnewline
 & $1^{3}D_{2^{++}(\bar{3}3)_{c}\left(\xi_{1},\xi_{2}\right)}$ & $\left(7085\right)$ & 7085 & 1\tabularnewline
\hline\hline
\end{tabular}
\end{center}
\end{table*}

%%%%%%%%%%%%%%%%%%%%%%%%%%%%%%%%%%%%%%%%

%%%%%%%%%%%%%%%%%%%%%%%%%%%%%%%%%%%%%%%%
\begin{table*}[htp]
\begin{center}
\caption{\label{mass of bbbb1D1} Predicted mass spectrum for the $1D$-wave $T_{(bb\bar{b}\bar{b})}$ states.}

\begin{tabular}{ccccc}
\hline\hline
$J^{PC}$ & Configuration & $\langle H\rangle$ (MeV) & Mass (MeV) & Eigenvector\tabularnewline
\hline
\multirow{2}{*}{$0^{+-}$} & $\begin{array}{l}
1^{3}P_{0^{+-}(6\bar{6})_{c}\left(\xi_{1}\xi_{3},\xi_{2}\xi_{3}\right)}\\
1^{3}P_{0^{+-}(\bar{3}3)_{c}\left(\xi_{1}\xi_{3},\xi_{2}\xi_{3}\right)}
\end{array}$ & $\left(\begin{array}{cc}
19693 & -116\\
-116 & 19775
\end{array}\right)$ & $\left(\begin{array}{c}
19611\\
19857
\end{array}\right)$ & $\left(\begin{array}{cc}
-0.82 & -0.58\\
0.58 & -0.82
\end{array}\right)$\tabularnewline
 & $1^{5}D_{0^{+-}(\bar{3}3)_{c}\left(\xi_{1},\xi_{2}\right)}$ & $\left(19884\right)$ & 19884 & 1\tabularnewline
\hline
\multirow{5}{*}{$0^{++}$} & $1^{1}S_{0^{++}(\bar{3}3)_{c}\left(\xi_{1}\xi_{2}\right)}$ & $\left(19677\right)$ & 19677 & 1\tabularnewline
 & $1^{1}S_{0^{++}(6\bar{6})_{c}\left(\xi_{1}\xi_{2}\right)}$ & $\left(19796\right)$ & 19796 & 1\tabularnewline
 & $\begin{array}{l}
1^{3}P_{0^{++}(6\bar{6})_{c}\left(\xi_{1}\xi_{3},\xi_{2}\xi_{3}\right)}\\
1^{3}P_{0^{++}(\bar{3}3)_{c}\left(\xi_{1}\xi_{3},\xi_{2}\xi_{3}\right)}
\end{array}$ & $\left(\begin{array}{cc}
19688 & 125\\
125 & 19773
\end{array}\right)$ & $\left(\begin{array}{c}
19598\\
19863
\end{array}\right)$ & $\left(\begin{array}{cc}
-0.81 & 0.58\\
0.58 & 0.81
\end{array}\right)$\tabularnewline
 & $1^{3}P_{0^{++}(6\bar{6})_{c}\left(\xi_{1}\xi_{2}\right)}$ & $\left(19884\right)$ & 19884 & 1\tabularnewline
 & $\begin{array}{l}
1^{5}D_{0^{++}(\bar{3}3)_{c}\left(\xi_{1},\xi_{2}\right)}\\
1^{5}D_{0^{++}(\bar{3}3)_{c}\left(\xi_{3}\right)}\\
1^{5}D_{0^{++}(6\bar{6})_{c}\left(\xi_{1}\xi_{2}\right)}
\end{array}$ & $\left(\begin{array}{ccc}
19873 & 59 & 22\\
59 & 19800 & -126\\
22 & -126 & 19847
\end{array}\right)$ & $\left(\begin{array}{c}
19885\\
19677\\
19958
\end{array}\right)$ & $\left(\begin{array}{ccc}
0.91 & 0.04 & 0.4\\
-0.29 & 0.75 & 0.59\\
0.28 & 0.66 & -0.7
\end{array}\right)$\tabularnewline
\hline\hline
\end{tabular}
\end{center}
\end{table*}

%%%%%%%%%%%%%%%%%%%%%%%%%%%%%%%%%%%%%%%%
\begin{table*}[htp]
\begin{center}
\caption{\label{mass of bbbb1D2} Predicted mass spectrum for the $1D$-wave $T_{(bb\bar{b}\bar{b})}$ states (Continued).}

\begin{tabular}{ccccc}
\hline\hline
$J^{PC}$ & Configuration & $\langle H\rangle$ (MeV) & Mass (MeV) & Eigenvector\tabularnewline
\hline
\multirow{8}{*}{$2^{++}$} & $1^{5}S_{2^{++}(6\bar{6})_{c}\left(\xi_{1}\xi_{2}\right)}$ & $\left(19814\right)$ & 19814 & 1\tabularnewline
 & $\begin{array}{l}
1^{3}P_{2^{++}(6\bar{6})_{c}\left(\xi_{1}\xi_{3},\xi_{2}\xi_{3}\right)}\\
1^{3}P_{2^{++}(\bar{3}3)_{c}\left(\xi_{1}\xi_{3},\xi_{2}\xi_{3}\right)}
\end{array}$ & $\left(\begin{array}{cc}
19691 & 129\\
129 & 19777
\end{array}\right)$ & $\left(\begin{array}{c}
19598\\
19870
\end{array}\right)$ & $\left(\begin{array}{cc}
-0.81 & 0.58\\
0.58 & 0.81
\end{array}\right)$\tabularnewline
 & $1^{3}P_{2^{++}(6\bar{6})_{c}\left(\xi_{1}\xi_{2}\right)}$ & $\left(19891\right)$ & 19891 & 1\tabularnewline
 & $\begin{array}{l}
1^{1}D_{2^{++}(6\bar{6})_{c}\left(\xi_{1},\xi_{2}\right)}\\
1^{1}D_{2^{++}(6\bar{6})_{c}\left(\xi_{3}\right)}\\
1^{1}D_{2^{++}(\bar{3}3)_{c}\left(\xi_{1}\otimes\xi_{2}\right)}
\end{array}$ & $\left(\begin{array}{ccc}
19773 & 156 & 29\\
156 & 19770 & -133\\
29 & -133 & 19697
\end{array}\right)$ & $\left(\begin{array}{c}
19952\\
19758\\
19530
\end{array}\right)$ & $\left(\begin{array}{ccc}
-0.59 & -0.74 & 0.32\\
0.64 & -0.2 & 0.74\\
0.48 & -0.64 & -0.59
\end{array}\right)$\tabularnewline
 & $\begin{array}{l}
1^{1}D_{2^{++}(\bar{3}3)_{c}\left(\xi_{1},\xi_{2}\right)}\\
1^{1}D_{2^{++}(\bar{3}3)_{c}\left(\xi_{3}\right)}\\
1^{1}D_{2^{++}(6\bar{6})_{c}\left(\xi_{1}\otimes\xi_{2}\right)}
\end{array}$ & $\left(\begin{array}{ccc}
19881 & 63 & 30\\
63 & 19806 & -135\\
30 & -135 & 19848
\end{array}\right)$ & $\left(\begin{array}{c}
19968\\
19669\\
19898
\end{array}\right)$ & $\left(\begin{array}{ccc}
0.25 & 0.67 & -0.7\\
-0.3 & 0.74 & 0.61\\
0.92 & 0.06 & 0.39
\end{array}\right)$\tabularnewline
 & $1^{3}D_{2^{++}(\bar{3}3)_{c}\left(\xi_{1},\xi_{2}\right)}$ & $\left(19894\right)$ & 19894 & 1\tabularnewline
\hline\hline
\end{tabular}
\end{center}
\end{table*}

%%%%%%%%%%%%%%%%%%%%%%%%%%%%%%%%%%%%%%%%
\begin{table*}[htp]
\begin{center}
\caption{\label{mass of cccc2Sa} The average contributions of each part of the Hamiltonian to the $2S$- and $3S$-wave $T_{(cc\bar{c}\bar{c})}$ configurations. $\langle T\rangle$ stands for the contribution of the kinetic energy term. $\langle V^{Lin}\rangle$ and $\langle V^{Coul}\rangle$ stand for the contributions from the linear confinement potential and Coulomb type potential, respectively. $\langle V^{SS}\rangle$, $\langle V^{T}\rangle$, and $\langle V^{LS}\rangle$ stand for the contributions from the spin-spin interaction term, the tensor potential term, and the spin-orbit interaction term, respectively.}
\scalebox{1.0}{
\begin{tabular}{cp{3cm}p{1cm}<{\centering}p{1cm}<{\centering}p{1cm}<{\centering}p{1cm}<{\centering}p{1cm}<{\centering}p{1cm}<{\centering}p{1cm}<{\centering} }
\hline\hline
$J^{PC}$ & Configuration & Mass &	~~~~$\langle T\rangle$~~~~ & ~~~~$\langle V^{Lin}\rangle$~~~~ &	 ~~~~$\langle V^{Coul}\rangle$~~~~ &	 ~~~~$\langle V^{SS}\rangle$~~~~	  \\	
\hline
$0^{+-}$&	$	2^{1}S_{0^{+-}(6\bar{6})_{c}\left(\xi_{1},\xi_{2}\right)}	$	&		7008	&	706	&	 899	&	-540	&	10.18					 \\
		&	$	2^{1}S_{0^{+-}(\bar{3}3)_{c}\left(\xi_{1},\xi_{2}\right)}	$	&		7005	&	776	&	 924	&	-629	&	2.75					 \\
\tabularnewline																					
		&	$	2^{1}S_{0^{++}(6\bar{6})_{c}\left(\xi_{1},\xi_{2}\right)}	$	&		6954	&	725	&	 883	&	-598	&	11.39					 \\
$0^{++}$&	$	2^{1}S_{0^{++}(\bar{3}3)_{c}\left(\xi_{1},\xi_{2}\right)}	$	&		7000	&	774	&	 919	&	-622	&	-3.79					 \\
		&	$	2^{1}S_{0^{++}(6\bar{6})_{c}\left(\xi_{3}\right)}	$	&		7183	&	757	&	1010	&	 -522	&	6.52					 \\
		&	$	2^{1}S_{0^{++}(\bar{3}3)_{c}\left(\xi_{3}\right)}	$	&		6930	&	761	&	876	&	 -642	&	2.56					\\
\tabularnewline																					
$1^{+-}$&	$	2^{3}S_{1^{+-}(\bar{3}3)_{c}\left(\xi_{1},\xi_{2}\right)}	$	&		7006	&	774	&	 920	&	-622	&	2.32					 \\
		&	$	2^{3}S_{1^{+-}(\bar{3}3)_{c}\left(\xi_{3}\right)}	$	&		6934	&	745	&	885	&	 -634	&	6.18					\\
\tabularnewline																					
$1^{++}$&	$	2^{3}S_{1^{++}(\bar{3}3)_{c}\left(\xi_{1},\xi_{2}\right)}	$	&		7009	&	773	&	 925	&	-628	&	6.88					 \\
\tabularnewline																					
$2^{+-}$&	$	2^{5}S_{2^{+-}(\bar{3}3)_{c}\left(\xi_{1},\xi_{2}\right)}	$	&		7017	&	762	&	 932	&	-624	&	14.99					 \\
\tabularnewline																					
$2^{++}$&	$	2^{5}S_{2^{++}(\bar{3}3)_{c}\left(\xi_{1},\xi_{2}\right)}	$	&		7018	&	753	&	 932	&	-613	&	13.94					 \\
		&	$	2^{5}S_{2^{++}(\bar{3}3)_{c}\left(\xi_{3}\right)}	$	&		6942	&	741	&	888	&	 -633	&	13.96					\\
\hline\tabularnewline
$0^{+-}$		&	$	3^{1}S_{0^{+-}(6\bar{6})_{c}\left(\xi_{1},\xi_{2}\right)}	$	&		7356	&	 761	&	1103	&	-449	&	8.75					 \\
		&	$	3^{1}S_{0^{+-}(\bar{3}3)_{c}\left(\xi_{1},\xi_{2}\right)}	$	&		7396	&	839	&	 1151	&	-528	&	1.9					 \\
\tabularnewline																					
		&	$	3^{1}S_{0^{++}(6\bar{6})_{c}\left(\xi_{1},\xi_{2}\right)}	$	&		7347	&	764	&	 1097	&	-455	&	9.02					 \\
$0^{++}$		&	$	3^{1}S_{0^{++}(\bar{3}3)_{c}\left(\xi_{1},\xi_{2}\right)}	$	&		7403	&	 838	&	1150	&	-518	&	0.53					 \\
		&	$	3^{1}S_{0^{++}(6\bar{6})_{c}\left(\xi_{3}\right)}	$	&		7720	&	855	&	1326	&	 -397	&	4.59					 \\
		&	$	3^{1}S_{0^{++}(\bar{3}3)_{c}\left(\xi_{3}\right)}	$	&		7241	&	815	&	1064	&	 -575	&	4.46					 \\
\tabularnewline																					
$1^{+-}$		&	$	3^{3}S_{1^{+-}(\bar{3}3)_{c}\left(\xi_{1},\xi_{2}\right)}	$	&		7406	&	 828	&	1157	&	-515	&	3.6					 \\
		&	$	3^{3}S_{1^{+-}(\bar{3}3)_{c}\left(\xi_{3}\right)}	$	&		7243	&	814	&	1065	&	 -575	&	7.06					 \\
\tabularnewline																					
$1^{++}$		&	$	3^{3}S_{1^{++}(\bar{3}3)_{c}\left(\xi_{1},\xi_{2}\right)}	$	&		7399	&	 838	&	1152	&	-528	&	4.88					 \\
\tabularnewline																					
$2^{+-}$		&	$	3^{5}S_{2^{+-}(\bar{3}3)_{c}\left(\xi_{1},\xi_{2}\right)}	$	&		7405	&	 835	&	1154	&	-527	&	10.81					 \\
\tabularnewline																					
$2^{++}$		&	$	3^{5}S_{2^{++}(\bar{3}3)_{c}\left(\xi_{1},\xi_{2}\right)}	$	&		7412	&	 825	&	1160	&	-514	&	9.67					 \\
		&	$	3^{5}S_{2^{++}(\bar{3}3)_{c}\left(\xi_{3}\right)}	$	&		7248	&	812	&	1067	&	 -575	&	12.3					 \\
\hline\hline
\end{tabular}}
\end{center}
\end{table*}

%%%%%%%%%%%%%%%%%%%%%%%%%%%%%%%%%%%%%%%%
\begin{table*}[htp]
\begin{center}
\caption{\label{mass of bbbb2Sa} The average contributions of each part of the Hamiltonian to the $2S$- and $3S$-wave $T_{(bb\bar{b}\bar{b})}$ configurations.}
\scalebox{1.0}{
\begin{tabular}{cp{3cm}p{1cm}<{\centering}p{1cm}<{\centering}p{1cm}<{\centering}p{1cm}<{\centering}p{1cm}<{\centering}p{1cm}<{\centering}p{1cm}<{\centering}cccccccccccc}
\hline\hline
$J^{PC}$ & Configuration & Mass &	$\langle T\rangle$ & $\langle V^{Lin}\rangle$ &	$\langle V^{Coul}\rangle$ &	$\langle V^{SS}\rangle$	 \\	
\hline
$0^{+-}$		&	$	2^{1}S_{0^{+-}(6\bar{6})_{c}\left(\xi_{1},\xi_{2}\right)}	$	&		19840	&	 627	&	529	&	-727	&	3.99					 \\
		&	$	2^{1}S_{0^{+-}(\bar{3}3)_{c}\left(\xi_{1},\xi_{2}\right)}	$	&		19790	&	701	&	 543	&	-864	&	2.28					 \\
\tabularnewline																					
		&	$	2^{1}S_{0^{++}(6\bar{6})_{c}\left(\xi_{1},\xi_{2}\right)}	$	&		19770	&	653	&	 515	&	-810	&	4.48					 \\
$0^{++}$		&	$	2^{1}S_{0^{++}(\bar{3}3)_{c}\left(\xi_{1},\xi_{2}\right)}	$	&		19797	&	 690	&	538	&	-837	&	-1.21					 \\
		&	$	2^{1}S_{0^{++}(6\bar{6})_{c}\left(\xi_{3}\right)}	$	&		19972	&	646	&	604	&	 -688	&	2.36					\\
		&	$	2^{1}S_{0^{++}(\bar{3}3)_{c}\left(\xi_{3}\right)}	$	&		19733	&	694	&	510	&	 -881	&	1.58					\\
\tabularnewline																					
$1^{+-}$		&	$	2^{3}S_{1^{+-}(\bar{3}3)_{c}\left(\xi_{1},\xi_{2}\right)}	$	&		19800	&	 690	&	538	&	-837	&	1.14					 \\
		&	$	2^{3}S_{1^{+-}(\bar{3}3)_{c}\left(\xi_{3}\right)}	$	&		19735	&	693	&	511	&	 -880	&	3.09					\\
\tabularnewline																					
$1^{++}$		&	$	2^{3}S_{1^{++}(\bar{3}3)_{c}\left(\xi_{1},\xi_{2}\right)}	$	&		19792	&	 700	&	543	&	-863	&	3.81					 \\
\tabularnewline																					
$2^{+-}$		&	$	2^{5}S_{2^{+-}(\bar{3}3)_{c}\left(\xi_{1},\xi_{2}\right)}	$	&		19795	&	 692	&	546	&	-859	&	6.83					 \\
\tabularnewline																					
$2^{++}$		&	$	2^{5}S_{2^{++}(\bar{3}3)_{c}\left(\xi_{1},\xi_{2}\right)}	$	&		19804	&	 671	&	545	&	-826	&	5.60					 \\
		&	$	2^{5}S_{2^{++}(\bar{3}3)_{c}\left(\xi_{3}\right)}	$	&		19738	&	692	&	512	&	 -880	&	6.11					\\
\hline\tabularnewline
$0^{+-}$		&	$	3^{1}S_{0^{+-}(6\bar{6})_{c}\left(\xi_{1},\xi_{2}\right)}	$	&		20128	&	 622	&	675	&	-580	&	3.07					 \\
		&	$	3^{1}S_{0^{+-}(\bar{3}3)_{c}\left(\xi_{1},\xi_{2}\right)}	$	&		20115	&	696	&	 701	&	-691	&	1.26					 \\
\tabularnewline																					
		&	$	3^{1}S_{0^{++}(6\bar{6})_{c}\left(\xi_{1},\xi_{2}\right)}	$	&		20120	&	627	&	 669	&	-587	&	3.19					 \\
$0^{++}$		&	$	3^{1}S_{0^{++}(\bar{3}3)_{c}\left(\xi_{1},\xi_{2}\right)}	$	&		20128	&	 688	&	703	&	-671	&	0.49					 \\
		&	$	3^{1}S_{0^{++}(6\bar{6})_{c}\left(\xi_{3}\right)}	$	&		20405	&	661	&	833	&	 -498	&	1.41					\\
		&	$	3^{1}S_{0^{++}(\bar{3}3)_{c}\left(\xi_{3}\right)}	$	&		19979	&	710	&	640	&	 -782	&	2.37					\\
\tabularnewline																					
$1^{+-}$		&	$	3^{3}S_{1^{+-}(\bar{3}3)_{c}\left(\xi_{1},\xi_{2}\right)}	$	&		20129	&	 687	&	703	&	-670	&	1.54					 \\
		&	$	3^{3}S_{1^{+-}(\bar{3}3)_{c}\left(\xi_{3}\right)}	$	&		19980	&	710	&	641	&	 -782	&	3.3					\\
\tabularnewline																					
$1^{++}$		&	$	3^{3}S_{1^{++}(\bar{3}3)_{c}\left(\xi_{1},\xi_{2}\right)}	$	&		20116	&	 696	&	702	&	-691	&	2.27					 \\
\tabularnewline																					
$2^{+-}$		&	$	3^{5}S_{2^{+-}(\bar{3}3)_{c}\left(\xi_{1},\xi_{2}\right)}	$	&		20118	&	 695	&	702	&	-691	&	4.28					 \\
\tabularnewline																					
$2^{++}$		&	$	3^{5}S_{2^{++}(\bar{3}3)_{c}\left(\xi_{1},\xi_{2}\right)}	$	&		20131	&	 685	&	704	&	-670	&	3.64					 \\
		&	$	3^{5}S_{2^{++}(\bar{3}3)_{c}\left(\xi_{3}\right)}	$	&		19982	&	709	&	641	&	 -782	&	5.19					\\
\hline\hline
\end{tabular}}
\end{center}
\end{table*}

%%%%%%%%%%%%%%%%%%%%%%%%%%%%%%%%%%%%%%%%
\begin{table*}[htp]
\begin{center}
\caption{\label{mass of cccc1D1a} The average contributions of each part of the Hamiltonian to the $1D$-wave $T_{(cc\bar{c}\bar{c})}$ configurations. $\langle T\rangle$ stands for the contribution of the kinetic energy term. $\langle V^{Lin}\rangle$ and $\langle V^{Coul}\rangle$ stand for the contributions from the linear confinement potential and Coulomb type potential, respectively. $\langle V^{SS}\rangle$, $\langle V^{T}\rangle$, and $\langle V^{LS}\rangle$ stand for the contributions from the spin-spin interaction term, the tensor potential term, and the spin-orbit interaction term, respectively.}
\begin{tabular}{ccccccccc}
\hline\hline
$J^{PC}$ & Configuration & ~~~~ Mass~~~~ &~~~~ $\langle T\rangle$ ~~~~& ~~~~ $\langle V^{Lin}\rangle$ ~~~~& ~~~~$\langle V^{Coul}\rangle$ ~~~~& ~~~~ $\langle V^{SS}\rangle$ ~~~~&~~~~ $\langle V^{T}\rangle$ ~~~~& ~~~~ $\langle V^{LS}\rangle$\tabularnewline
\hline
\multirow{3}{*}{$0^{+-}$} & $1^{3}P_{0^{+-}(6\bar{6})_{c}\left(\xi_{1}\xi_{3},\xi_{2}\xi_{3}\right)}$ & 6868 & 732 & 837 & -636 & 6.18 & 0 & -2.28\tabularnewline
 & $1^{3}P_{0^{+-}(\bar{3}3)_{c}\left(\xi_{1}\xi_{3},\xi_{2}\xi_{3}\right)}$ & 6948 & 749 & 876 & -601 & -2.66 & 0 & -4.73\tabularnewline
 & $1^{5}D_{0^{+-}(\bar{3}3)_{c}\left(\xi_{1},\xi_{2}\right)}$ & 7054 & 809 & 924 & -572 & 9.25 & -5.91 & -43.52\tabularnewline
 &  &  &  &  &  &  &  & \tabularnewline
\multirow{8}{*}{$0^{++}$} & $1^{1}S_{0^{++}(\bar{3}3)_{c}\left(\xi_{1}\xi_{2}\right)}$ & 6838 & 769 & 818 & -657 & -24.14 & 0 & 0\tabularnewline
 & $1^{1}S_{0^{++}(6\bar{6})_{c}\left(\xi_{1}\xi_{2}\right)}$ & 6957 & 788 & 877 & -603 & -37.55 & 0 & 0\tabularnewline
 & $1^{3}P_{0^{++}(6\bar{6})_{c}\left(\xi_{1}\xi_{3},\xi_{2}\xi_{3}\right)}$ & 6857 & 745 & 830 & -642 & -2.02 & -2.35 & -2.35\tabularnewline
 & $1^{3}P_{0^{++}(\bar{3}3)_{c}\left(\xi_{1}\xi_{3},\xi_{2}\xi_{3}\right)}$ & 6944 & 754 & 873 & -603 & -6.08 & -0.96 & -4.79\tabularnewline
 & $1^{3}P_{0^{++}(6\bar{6})_{c}\left(\xi_{1}\xi_{2}\right)}$ & 7053 & 775 & 925 & -553 & -15.43 & 2.49 & -12.47\tabularnewline
 & $1^{5}D_{0^{++}(\bar{3}3)_{c}\left(\xi_{1},\xi_{2}\right)}$ & 7051 & 798 & 927 & -577 & 9.84 & -4.46 & -35.39\tabularnewline
 & $1^{5}D_{0^{++}(\bar{3}3)_{c}\left(\xi_{3}\right)}$ & 6968 & 783 & 883 & -603 & 10.84 & -4.59 & -32.12\tabularnewline
 & $1^{5}D_{0^{++}(6\bar{6})_{c}\left(\xi_{1}\xi_{2}\right)}$ & 7013 & 802 & 902 & -585 & 9.78 & -5.96 & -41.73\tabularnewline
 &  &  &  &  &  &  &  & \tabularnewline
\multirow{14}{*}{$1^{+-}$} & $1^{3}S_{1^{+-}(6\bar{6})_{c}\left(\xi_{1}\xi_{3},\xi_{2}\xi_{3}\right)}$ & 7002 & 739 & 908 & -584 & 7.19 & 0 & 0\tabularnewline
 & $1^{3}S_{1^{+-}(\bar{3}3)_{c}\left(\xi_{1}\xi_{3},\xi_{2}\xi_{3}\right)}$ & 7007 & 743 & 906 & -571 & -2.36 & 0 & 0\tabularnewline
 & $1^{3}S_{1^{+-}(6\bar{6})_{c}\left(\xi_{1}\xi_{2}\right)}$ & 6973 & 768 & 888 & -595 & -20.14 & 0 & 0\tabularnewline
 & $1^{1}P_{1^{+-}(\bar{3}3)_{c}\left(\xi_{1}\xi_{2}\right)}$ & 6871 & 767 & 834 & -637 & -23.89 & 0 & 0\tabularnewline
 & $1^{1}P_{1^{+-}(6\bar{6})_{c}\left(\xi_{1}\xi_{2}\right)}$ & 7052 & 776 & 924 & -554 & -26.86 & 0 & 0\tabularnewline
 & $1^{3}P_{1^{+-}(6\bar{6})_{c}\left(\xi_{1}\xi_{3},\xi_{2}\xi_{3}\right)}$ & 6867 & 733 & 836 & -637 & 6.2 & -0.86 & -2.58\tabularnewline
 & $1^{3}P_{1^{+-}(\bar{3}3)_{c}\left(\xi_{1}\xi_{3},\xi_{2}\xi_{3}\right)}$ & 6946 & 753 & 874 & -603 & -2.68 & -1.79 & -5.37\tabularnewline
 & $1^{5}P_{1^{+-}(6\bar{6})_{c}\left(\xi_{1}\xi_{2}\right)}$ & 7069 & 754 & 937 & -545 & 7.04 & 0.6 & -16.07\tabularnewline
 & $1^{3}D_{1^{+-}(\bar{3}3)_{c}\left(\xi_{1},\xi_{2}\right)}$ & 7062 & 781 & 937 & -570 & -1.23 & -0.13 & -17.01\tabularnewline
 & $1^{3}D_{1^{+-}(\bar{3}3)_{c}\left(\xi_{3}\right)}$ & 6978 & 767 & 891 & -597 & -1.51 & 1.23 & -15.46\tabularnewline
 & $1^{3}D_{1^{+-}(6\bar{6})_{c}\left(\xi_{1}\xi_{3},\xi_{2}\xi_{3}\right)}$ & 7092 & 782 & 953 & -555 & 1.73 & -0.06 & -19.86\tabularnewline
 & $1^{3}D_{1^{+-}(\bar{3}3)_{c}\left(\xi_{1}\xi_{3},\xi_{2}\xi_{3}\right)}$ & 7026 & 786 & 912 & -575 & -4.9 & -2.18 & -21.88\tabularnewline
 & $1^{3}D_{1^{+-}(6\bar{6})_{c}\left(\xi_{1}\xi_{2}\right)}$ & 7008 & 804 & 901 & -586 & -18.28 & -2.98 & -20.89\tabularnewline
 & $1^{5}D_{1^{+-}(\bar{3}3)_{c}\left(\xi_{1},\xi_{2}\right)}$ & 7064 & 794 & 933 & -566 & 8.98 & -2.86 & -35.06\tabularnewline
 &  &  &  &  &  &  &  & \tabularnewline
\multirow{11}{*}{$1^{++}$} & $1^{3}S_{1^{++}(6\bar{6})_{c}\left(\xi_{1}\xi_{3},\xi_{2}\xi_{3}\right)}$ & 6991 & 747 & 903 & -588 & -3.32 & 0 & 0\tabularnewline
 & $1^{3}S_{1^{++}(\bar{3}3)_{c}\left(\xi_{1}\xi_{3},\xi_{2}\xi_{3}\right)}$ & 7003 & 746 & 903 & -572 & -6.56 & 0 & 0\tabularnewline
 & $1^{3}P_{1^{++}(6\bar{6})_{c}\left(\xi_{1}\xi_{3},\xi_{2}\xi_{3}\right)}$ & 6861 & 739 & 833 & -640 & -2 & 2.03 & -2.61\tabularnewline
 & $1^{3}P_{1^{++}(\bar{3}3)_{c}\left(\xi_{1}\xi_{3},\xi_{2}\xi_{3}\right)}$ & 6944 & 755 & 873 & -603 & -6.09 & -0.6 & -5.39\tabularnewline
 & $1^{3}P_{1^{++}(6\bar{6})_{c}\left(\xi_{1}\xi_{2}\right)}$ & 7051 & 779 & 923 & -555 & -15.54 & -3.14 & -9.43\tabularnewline
 & $1^{3}D_{1^{++}(\bar{3}3)_{c}\left(\xi_{1},\xi_{2}\right)}$ & 7070 & 783 & 939 & -562 & -0.47 & -0.76 & -20.53\tabularnewline
 & $1^{3}D_{1^{++}(6\bar{6})_{c}\left(\xi_{1}\xi_{3},\xi_{2}\xi_{3}\right)}$ & 7096 & 777 & 956 & -554 & 2.85 & 1.81 & -19.64\tabularnewline
 & $1^{3}D_{1^{++}(\bar{3}3)_{c}\left(\xi_{1}\xi_{3},\xi_{2}\xi_{3}\right)}$ & 7028 & 784 & 913 & -574 & -4.41 & -1.4 & -21.78\tabularnewline
 & $1^{5}D_{1^{++}(\bar{3}3)_{c}\left(\xi_{1},\xi_{2}\right)}$ & 7059 & 786 & 934 & -572 & 9.62 & -2.17 & -28.72\tabularnewline
 & $1^{5}D_{1^{++}(\bar{3}3)_{c}\left(\xi_{3}\right)}$ & 6976 & 772 & 889 & -599 & 10.6 & -2.24 & -26.09\tabularnewline
 & $1^{5}D_{1^{++}(6\bar{6})_{c}\left(\xi_{1}\xi_{2}\right)}$ & 7022 & 787 & 911 & -580 & 9.49 & -2.88 & -33.63\tabularnewline
 &  &  &  &  &  &  &  & \tabularnewline
\multirow{11}{*}{$2^{+-}$} & $1^{3}P_{2^{+-}(6\bar{6})_{c}\left(\xi_{1}\xi_{3},\xi_{2}\xi_{3}\right)}$ & 6873 & 725 & 841 & -633 & 6.1 & 0.17 & 2.53\tabularnewline
 & $1^{3}P_{2^{+-}(\bar{3}3)_{c}\left(\xi_{1}\xi_{3},\xi_{2}\xi_{3}\right)}$ & 6959 & 735 & 884 & -595 & -2.57 & 0.34 & 5.15\tabularnewline
 & $1^{5}P_{2^{+-}(6\bar{6})_{c}\left(\xi_{1}\xi_{2}\right)}$ & 7072 & 751 & 939 & -544 & 6.99 & -4.13 & -8.86\tabularnewline
 & $1^{1}D_{2^{+-}(6\bar{6})_{c}\left(\xi_{1},\xi_{2}\right)}$ & 6985 & 727 & 895 & -575 & 5.62 & 0 & 0\tabularnewline
 & $1^{1}D_{2^{+-}(\bar{3}3)_{c}\left(\xi_{1},\xi_{2}\right)}$ & 7087 & 760 & 953 & -553 & -4.84 & 0 & 0\tabularnewline
 & $1^{3}D_{2^{+-}(\bar{3}3)_{c}\left(\xi_{1},\xi_{2}\right)}$ & 7073 & 765 & 947 & -564 & -1.19 & 0.12 & -5.47\tabularnewline
 & $1^{3}D_{2^{+-}(\bar{3}3)_{c}\left(\xi_{3}\right)}$ & 6985 & 755 & 898 & -592 & -1.48 & -1.2 & -5.02\tabularnewline
 & $1^{3}D_{2^{+-}(6\bar{6})_{c}\left(\xi_{1}\xi_{3},\xi_{2}\xi_{3}\right)}$ & 7106 & 763 & 964 & -549 & 1.67 & 0.06 & -6.35\tabularnewline
 & $1^{3}D_{2^{+-}(\bar{3}3)_{c}\left(\xi_{1}\xi_{3},\xi_{2}\xi_{3}\right)}$ & 7045 & 760 & 927 & -565 & -4.63 & 2.05 & -6.87\tabularnewline
 & $1^{3}D_{2^{+-}(6\bar{6})_{c}\left(\xi_{1}\xi_{2}\right)}$ & 7028 & 774 & 917 & -575 & -17.3 & 2.8 & -6.53\tabularnewline
 & $1^{5}D_{2^{+-}(\bar{3}3)_{c}\left(\xi_{1},\xi_{2}\right)}$ & 7082 & 768 & 948 & -556 & 8.53 & 1.16 & -19.86\tabularnewline
\hline\hline
\end{tabular}
\end{center}
\end{table*}

%%%%%%%%%%%%%%%%%%%%%%%%%%%%%%%%%%%%%%%%
\begin{table*}[htp]
\begin{center}
\caption{\label{mass of cccc1D2a} The average contributions of each part of the Hamiltonian to the $1D$-wave $T_{(cc\bar{c}\bar{c})}$ configurations (Continued).}
\begin{tabular}{ccccccccc}
\hline\hline
$J^{PC}$ & Configuration & ~~~~ Mass~~~~ &~~~~ $\langle T\rangle$ ~~~~& ~~~~ $\langle V^{Lin}\rangle$ ~~~~& ~~~~$\langle V^{Coul}\rangle$ ~~~~& ~~~~ $\langle V^{SS}\rangle$ ~~~~&~~~~ $\langle V^{T}\rangle$ ~~~~& ~~~~ $\langle V^{LS}\rangle$\tabularnewline
\hline
\multirow{16}{*}{$2^{++}$} & $1^{5}S_{2^{++}(6\bar{6})_{c}\left(\xi_{1}\xi_{2}\right)}$ & 7004 & 731 & 911 & -581 & 11.30 & 0 & 0\tabularnewline
 & $1^{3}P_{2^{++}(6\bar{6})_{c}\left(\xi_{1}\xi_{3},\xi_{2}\xi_{3}\right)}$ & 6864 & 735 & 835 & -638 & -1.99 & -0.4 & 2.58\tabularnewline
 & $1^{3}P_{2^{++}(\bar{3}3)_{c}\left(\xi_{1}\xi_{3},\xi_{2}\xi_{3}\right)}$ & 6955 & 739 & 882 & -597 & -5.89 & 0.12 & 5.19\tabularnewline
 & $1^{3}P_{2^{++}(6\bar{6})_{c}\left(\xi_{1}\xi_{2}\right)}$ & 7073 & 748 & 941 & -543 & -14.59 & 0.59 & 8.79\tabularnewline
 & $1^{1}D_{2^{++}(6\bar{6})_{c}\left(\xi_{1},\xi_{2}\right)}$ & 6955 & 733 & 883 & -598 & 5.68 & 0 & 0\tabularnewline
 & $1^{1}D_{2^{++}(\bar{3}3)_{c}\left(\xi_{1},\xi_{2}\right)}$ & 7073 & 764 & 947 & -563 & -6.38 & 0 & 0\tabularnewline
 & $1^{1}D_{2^{++}(6\bar{6})_{c}\left(\xi_{3}\right)}$ & 6964 & 744 & 894 & -613 & 6.43 & 0 & 0\tabularnewline
 & $1^{1}D_{2^{++}(\bar{3}3)_{c}\left(\xi_{3}\right)}$ & 6986 & 754 & 899 & -591 & -7.33 & 0 & 0\tabularnewline
 & $1^{1}D_{2^{++}(\bar{3}3)_{c}\left(\xi_{1}\xi_{2}\right)}$ & 6859 & 769 & 829 & -647 & -24.01 & 0 & 0\tabularnewline
 & $1^{1}D_{2^{++}(6\bar{6})_{c}\left(\xi_{1}\xi_{2}\right)}$ & 7018 & 787 & 910 & -579 & -31.28 & 0 & 0\tabularnewline
 & $1^{3}D_{2^{++}(\bar{3}3)_{c}\left(\xi_{1},\xi_{2}\right)}$ & 7085 & 762 & 952 & -554 & -0.44 & 0.72 & -6.52\tabularnewline
 & $1^{3}D_{2^{++}(6\bar{6})_{c}\left(\xi_{1}\xi_{3},\xi_{2}\xi_{3}\right)}$ & 7105 & 764 & 963 & -549 & 2.77 & -1.76 & -6.36\tabularnewline
 & $1^{3}D_{2^{++}(\bar{3}3)_{c}\left(\xi_{1}\xi_{3},\xi_{2}\xi_{3}\right)}$ & 7045 & 760 & 927 & -565 & -4.2 & 1.33 & -6.87\tabularnewline
 & $1^{5}D_{2^{++}(\bar{3}3)_{c}\left(\xi_{1},\xi_{2}\right)}$ & 7073 & 766 & 946 & -564 & 9.24 & 0.89 & -16.46\tabularnewline
 & $1^{5}D_{2^{++}(\bar{3}3)_{c}\left(\xi_{3}\right)}$ & 6989 & 752 & 900 & -591 & 10.19 & 0.92 & -14.97\tabularnewline
 & $1^{5}D_{2^{++}(6\bar{6})_{c}\left(\xi_{1}\xi_{2}\right)}$ & 7040 & 761 & 925 & -570 & 9.03 & 1.17 & -19.07\tabularnewline
 &  &  &  &  &  &  &  & \tabularnewline
\multirow{7}{*}{$3^{+-}$} & $1^{5}P_{3^{+-}(6\bar{6})_{c}\left(\xi_{1}\xi_{2}\right)}$ & 7102 & 711 & 964 & -529 & 6.38 & 1.08 & 16.13\tabularnewline
 & $1^{3}D_{3^{+-}(\bar{3}3)_{c}\left(\xi_{1},\xi_{2}\right)}$ & 7089 & 743 & 960 & -555 & -1.14 & -0.03 & 10.41\tabularnewline
 & $1^{3}D_{3^{+-}(\bar{3}3)_{c}\left(\xi_{3}\right)}$ & 7002 & 733 & 911 & -583 & -1.43 & 0.32 & 9.54\tabularnewline
 & $1^{3}D_{3^{+-}(6\bar{6})_{c}\left(\xi_{1}\xi_{3},\xi_{2}\xi_{3}\right)}$ & 7124 & 739 & 979 & -539 & 1.6 & -0.02 & 12.01\tabularnewline
 & $1^{3}D_{3^{+-}(\bar{3}3)_{c}\left(\xi_{1}\xi_{3},\xi_{2}\xi_{3}\right)}$ & 7062 & 736 & 942 & -555 & -4.4 & -0.55 & 13\tabularnewline
 & $1^{3}D_{3^{+-}(6\bar{6})_{c}\left(\xi_{1}\xi_{2}\right)}$ & 7043 & 752 & 930 & -566 & -16.57 & -0.76 & 12.43\tabularnewline
 & $1^{5}D_{3^{+-}(\bar{3}3)_{c}\left(\xi_{1},\xi_{2}\right)}$ & 7103 & 740 & 965 & -545 & 8.04 & 2.89 & 0\tabularnewline
 &  &  &  &  &  &  &  & \tabularnewline
\multirow{6}{*}{$3^{++}$} & $1^{3}D_{3^{++}(\bar{3}3)_{c}\left(\xi_{1},\xi_{2}\right)}$ & 7103 & 738 & 967 & -545 & -0.4 & -0.2 & 12.34\tabularnewline
 & $1^{3}D_{3^{++}(6\bar{6})_{c}\left(\xi_{1}\xi_{3},\xi_{2}\xi_{3}\right)}$ & 7126 & 736 & 981 & -538 & 2.59 & 0.47 & 11.94\tabularnewline
 & $1^{3}D_{3^{++}(\bar{3}3)_{c}\left(\xi_{1}\xi_{3},\xi_{2}\xi_{3}\right)}$ & 7063 & 735 & 942 & -555 & -3.99 & -0.36 & 12.97\tabularnewline
 & $1^{5}D_{3^{++}(\bar{3}3)_{c}\left(\xi_{1},\xi_{2}\right)}$ & 7091 & 742 & 961 & -555 & 8.81 & 2.24 & 0\tabularnewline
 & $1^{5}D_{3^{++}(\bar{3}3)_{c}\left(\xi_{3}\right)}$ & 7005 & 730 & 913 & -582 & 9.74 & 2.32 & 0\tabularnewline
 & $1^{5}D_{3^{++}(6\bar{6})_{c}\left(\xi_{1}\xi_{2}\right)}$ & 7060 & 734 & 942 & -559 & 8.52 & 2.92 & 0\tabularnewline
 &  &  &  &  &  &  &  & \tabularnewline
\multirow{1}{*}{$4^{+-}$} & $1^{5}D_{4^{+-}(\bar{3}3)_{c}\left(\xi_{1},\xi_{2}\right)}$ & 7122 & 715 & 982 & -536 & 7.62 & -1.36 & 23.36\tabularnewline
 &  &  &  &  &  &  &  & \tabularnewline
\multirow{3}{*}{$4^{++}$} & $1^{5}D_{4^{++}(\bar{3}3)_{c}\left(\xi_{1},\xi_{2}\right)}$ & 7108 & 720 & 975 & -546 & 8.42 & -1.07 & 19.74\tabularnewline
 & $1^{5}D_{4^{++}(\bar{3}3)_{c}\left(\xi_{3}\right)}$ & 7020 & 710 & 925 & -573 & 9.34 & -1.11 & 18.09\tabularnewline
 & $1^{5}D_{4^{++}(6\bar{6})_{c}\left(\xi_{1}\xi_{2}\right)}$ & 7079 & 710 & 957 & -549 & 8.09 & -1.38 & 22.5\tabularnewline
\hline\hline
\end{tabular}
\end{center}
\end{table*}

%%%%%%%%%%%%%%%%%%%%%%%%%%%%%%%%%%%%%%%%
\begin{table*}[htp]
\begin{center}
\caption{\label{mass of bbbb1D1a} The average contributions of each part of the Hamiltonian to the $1D$-wave $T_{(bb\bar{b}\bar{b})}$ configurations.}
\begin{tabular}{ccccccccc}
\hline\hline
$J^{PC}$ & Configuration & ~~~~ Mass~~~~ &~~~~ $\langle T\rangle$ ~~~~& ~~~~ $\langle V^{Lin}\rangle$ ~~~~& ~~~~$\langle V^{Coul}\rangle$ ~~~~& ~~~~ $\langle V^{SS}\rangle$ ~~~~&~~~~ $\langle V^{T}\rangle$ ~~~~& ~~~~ $\langle V^{LS}\rangle$\tabularnewline
\hline
\multirow{3}{*}{$0^{+-}$} & $1^{3}P_{0^{+-}(6\bar{6})_{c}\left(\xi_{1}\xi_{3},\xi_{2}\xi_{3}\right)}$ & 19693 & 677 & 482 & -876 & 2.6 & 0 & -0.91\tabularnewline
 & $1^{3}P_{0^{+-}(\bar{3}3)_{c}\left(\xi_{1}\xi_{3},\xi_{2}\xi_{3}\right)}$ & 19775 & 667 & 514 & -812 & -0.98 & 0 & -1.77\tabularnewline
 & $1^{5}D_{0^{+-}(\bar{3}3)_{c}\left(\xi_{1},\xi_{2}\right)}$ & 19884 & 674 & 561 & -746 & 3.24 & -1.98 & -14.56\tabularnewline
 &  &  &  &  &  &  &  & \tabularnewline
\multirow{8}{*}{$0^{++}$} & $1^{1}S_{0^{++}(\bar{3}3)_{c}\left(\xi_{1}\xi_{2}\right)}$ & 19677 & 697 & 475 & -893 & -9.68 & 0 & 0\tabularnewline
 & $1^{1}S_{0^{++}(6\bar{6})_{c}\left(\xi_{1}\xi_{2}\right)}$ & 19796 & 681 & 523 & -801 & -13.94 & 0 & 0\tabularnewline
 & $1^{3}P_{0^{++}(6\bar{6})_{c}\left(\xi_{1}\xi_{3},\xi_{2}\xi_{3}\right)}$ & 19688 & 684 & 480 & -881 & -0.91 & -0.92 & -0.92\tabularnewline
 & $1^{3}P_{0^{++}(\bar{3}3)_{c}\left(\xi_{1}\xi_{3},\xi_{2}\xi_{3}\right)}$ & 19773 & 670 & 513 & -813 & -2.37 & -0.36 & -1.78\tabularnewline
 & $1^{3}P_{0^{++}(6\bar{6})_{c}\left(\xi_{1}\xi_{2}\right)}$ & 19884 & 655 & 557 & -728 & -5.53 & 0.86 & -4.3\tabularnewline
 & $1^{5}D_{0^{++}(\bar{3}3)_{c}\left(\xi_{1},\xi_{2}\right)}$ & 19873 & 674 & 559 & -757 & 3.56 & -1.53 & -12.13\tabularnewline
 & $1^{5}D_{0^{++}(\bar{3}3)_{c}\left(\xi_{3}\right)}$ & 19800 & 678 & 525 & -803 & 4.05 & -1.64 & -11.46\tabularnewline
 & $1^{5}D_{0^{++}(6\bar{6})_{c}\left(\xi_{1}\xi_{2}\right)}$ & 19847 & 676 & 544 & -768 & 3.49 & -2.04 & -14.25\tabularnewline
 &  &  &  &  &  &  &  & \tabularnewline
\multirow{14}{*}{$1^{+-}$} & $1^{3}S_{1^{+-}(6\bar{6})_{c}\left(\xi_{1}\xi_{3},\xi_{2}\xi_{3}\right)}$ & 19812 & 662 & 535 & -796 & 2.97 & 0 & 0\tabularnewline
 & $1^{3}S_{1^{+-}(\bar{3}3)_{c}\left(\xi_{1}\xi_{3},\xi_{2}\xi_{3}\right)}$ & 19828 & 653 & 536 & -768 & -0.73 & 0 & 0\tabularnewline
 & $1^{3}S_{1^{+-}(6\bar{6})_{c}\left(\xi_{1}\xi_{2}\right)}$ & 19802 & 671 & 526 & -796 & -7.59 & 0 & 0\tabularnewline
 & $1^{1}P_{1^{+-}(\bar{3}3)_{c}\left(\xi_{1}\xi_{2}\right)}$ & 19711 & 689 & 488 & -864 & -9.56 & 0 & 0\tabularnewline
 & $1^{1}P_{1^{+-}(6\bar{6})_{c}\left(\xi_{1}\xi_{2}\right)}$ & 19883 & 656 & 557 & -728 & -9.59 & 0 & 0\tabularnewline
 & $1^{3}P_{1^{+-}(6\bar{6})_{c}\left(\xi_{1}\xi_{3},\xi_{2}\xi_{3}\right)}$ & 19692 & 678 & 482 & -877 & 2.61 & -0.34 & -1.02\tabularnewline
 & $1^{3}P_{1^{+-}(\bar{3}3)_{c}\left(\xi_{1}\xi_{3},\xi_{2}\xi_{3}\right)}$ & 19774 & 669 & 514 & -813 & -0.98 & -0.67 & -2\tabularnewline
 & $1^{5}P_{1^{+-}(6\bar{6})_{c}\left(\xi_{1}\xi_{2}\right)}$ & 19890 & 646 & 561 & -722 & 2.51 & 0.21 & -5.67\tabularnewline
 & $1^{3}D_{1^{+-}(\bar{3}3)_{c}\left(\xi_{1},\xi_{2}\right)}$ & 19877 & 667 & 562 & -753 & -0.44 & -0.04 & -5.96\tabularnewline
 & $1^{3}D_{1^{+-}(\bar{3}3)_{c}\left(\xi_{3}\right)}$ & 19803 & 672 & 528 & -799 & -0.63 & 0.45 & -5.64\tabularnewline
 & $1^{3}D_{1^{+-}(6\bar{6})_{c}\left(\xi_{1}\xi_{3},\xi_{2}\xi_{3}\right)}$ & 19906 & 662 & 573 & -731 & 0.66 & -0.02 & -6.86\tabularnewline
 & $1^{3}D_{1^{+-}(\bar{3}3)_{c}\left(\xi_{1}\xi_{3},\xi_{2}\xi_{3}\right)}$ & 19856 & 668 & 548 & -758 & -1.73 & -0.75 & -7.58\tabularnewline
 & $1^{3}D_{1^{+-}(6\bar{6})_{c}\left(\xi_{1}\xi_{2}\right)}$ & 19845 & 678 & 543 & -769 & -6.61 & -1.02 & -7.15\tabularnewline
 & $1^{5}D_{1^{+-}(\bar{3}3)_{c}\left(\xi_{1},\xi_{2}\right)}$ & 19887 & 668 & 563 & -742 & 3.2 & -0.97 & -11.95\tabularnewline
 &  &  &  &  &  &  &  & \tabularnewline
\multirow{11}{*}{$1^{++}$} & $1^{3}S_{1^{++}(6\bar{6})_{c}\left(\xi_{1}\xi_{3},\xi_{2}\xi_{3}\right)}$ & 19807 & 667 & 533 & -799 & -1.48 & 0 & 0\tabularnewline
 & $1^{3}S_{1^{++}(\bar{3}3)_{c}\left(\xi_{1}\xi_{3},\xi_{2}\xi_{3}\right)}$ & 19827 & 655 & 535 & -770 & -2.46 & 0 & 0\tabularnewline
 & $1^{3}P_{1^{++}(6\bar{6})_{c}\left(\xi_{1}\xi_{3},\xi_{2}\xi_{3}\right)}$ & 19690 & 681 & 481 & -879 & -0.91 & 0.8 & -1.03\tabularnewline
 & $1^{3}P_{1^{++}(\bar{3}3)_{c}\left(\xi_{1}\xi_{3},\xi_{2}\xi_{3}\right)}$ & 19773 & 670 & 513 & -813 & -2.37 & -0.22 & -2.01\tabularnewline
 & $1^{3}P_{1^{++}(6\bar{6})_{c}\left(\xi_{1}\xi_{2}\right)}$ & 19883 & 657 & 556 & -728 & -5.55 & -1.08 & -3.24\tabularnewline
 & $1^{3}D_{1^{++}(\bar{3}3)_{c}\left(\xi_{1},\xi_{2}\right)}$ & 19889 & 664 & 565 & -740 & -0.13 & -0.26 & -7.1\tabularnewline
 & $1^{3}D_{1^{++}(6\bar{6})_{c}\left(\xi_{1}\xi_{3},\xi_{2}\xi_{3}\right)}$ & 19907 & 660 & 574 & -730 & 0.97 & 0.63 & -6.83\tabularnewline
 & $1^{3}D_{1^{++}(\bar{3}3)_{c}\left(\xi_{1}\xi_{3},\xi_{2}\xi_{3}\right)}$ & 19856 & 667 & 548 & -757 & -1.6 & -0.49 & -7.57\tabularnewline
 & $1^{5}D_{1^{++}(\bar{3}3)_{c}\left(\xi_{1},\xi_{2}\right)}$ & 19876 & 669 & 561 & -754 & 3.52 & -0.75 & -9.98\tabularnewline
 & $1^{5}D_{1^{++}(\bar{3}3)_{c}\left(\xi_{3}\right)}$ & 19802 & 673 & 527 & -800 & 4.01 & -0.81 & -9.44\tabularnewline
 & $1^{5}D_{1^{++}(6\bar{6})_{c}\left(\xi_{1}\xi_{2}\right)}$ & 19850 & 670 & 546 & -765 & 3.44 & -1 & -11.69\tabularnewline
 &  &  &  &  &  &  &  & \tabularnewline
\multirow{11}{*}{$2^{+-}$} & $1^{3}P_{2^{+-}(6\bar{6})_{c}\left(\xi_{1}\xi_{3},\xi_{2}\xi_{3}\right)}$ & 19695 & 674 & 483 & -874 & 2.58 & 0.07 & 1.01\tabularnewline
 & $1^{3}P_{2^{+-}(\bar{3}3)_{c}\left(\xi_{1}\xi_{3},\xi_{2}\xi_{3}\right)}$ & 19779 & 660 & 517 & -807 & -0.96 & 0.13 & 1.96\tabularnewline
 & $1^{5}P_{2^{+-}(6\bar{6})_{c}\left(\xi_{1}\xi_{2}\right)}$ & 19891 & 645 & 561 & -721 & 2.5 & -1.46 & -3.14\tabularnewline
 & $1^{1}D_{2^{+-}(6\bar{6})_{c}\left(\xi_{1},\xi_{2}\right)}$ & 19807 & 648 & 525 & -777 & 2.21 & 0 & 0\tabularnewline
 & $1^{1}D_{2^{+-}(\bar{3}3)_{c}\left(\xi_{1},\xi_{2}\right)}$ & 19895 & 654 & 569 & -734 & -1.73 & 0 & 0\tabularnewline
 & $1^{3}D_{2^{+-}(\bar{3}3)_{c}\left(\xi_{1},\xi_{2}\right)}$ & 19881 & 660 & 564 & -749 & -0.43 & 0.04 & -1.95\tabularnewline
 & $1^{3}D_{2^{+-}(\bar{3}3)_{c}\left(\xi_{3}\right)}$ & 19806 & 667 & 529 & -796 & -0.62 & -0.44 & -1.86\tabularnewline
 & $1^{3}D_{2^{+-}(6\bar{6})_{c}\left(\xi_{1}\xi_{3},\xi_{2}\xi_{3}\right)}$ & 19910 & 654 & 576 & -727 & 0.65 & 0.02 & -2.24\tabularnewline
 & $1^{3}D_{2^{+-}(\bar{3}3)_{c}\left(\xi_{1}\xi_{3},\xi_{2}\xi_{3}\right)}$ & 19863 & 657 & 552 & -751 & -1.69 & 0.73 & -2.46\tabularnewline
 & $1^{3}D_{2^{+-}(6\bar{6})_{c}\left(\xi_{1}\xi_{2}\right)}$ & 19852 & 666 & 548 & -763 & -6.44 & 0.99 & -2.31\tabularnewline
 & $1^{5}D_{2^{+-}(\bar{3}3)_{c}\left(\xi_{1},\xi_{2}\right)}$ & 19893 & 657 & 567 & -736 & 3.12 & 0.41 & -6.98\tabularnewline
\hline\hline
\end{tabular}
\end{center}
\end{table*}

%%%%%%%%%%%%%%%%%%%%%%%%%%%%%%%%%%%%%%%%
\begin{table*}[htp]
\begin{center}
\caption{\label{mass of bbbb1D2a} The average contributions of each part of the Hamiltonian to the $1D$-wave $T_{(bb\bar{b}\bar{b})}$ configurations (Continued).}
\begin{tabular}{ccccccccc}
\hline\hline
$J^{PC}$ & Configuration & ~~~~ Mass~~~~ &~~~~ $\langle T\rangle$ ~~~~& ~~~~ $\langle V^{Lin}\rangle$ ~~~~& ~~~~$\langle V^{Coul}\rangle$ ~~~~& ~~~~ $\langle V^{SS}\rangle$ ~~~~&~~~~ $\langle V^{T}\rangle$ ~~~~& ~~~~ $\langle V^{LS}\rangle$\tabularnewline
\hline
\multirow{16}{*}{$2^{++}$} & $1^{5}S_{2^{++}(6\bar{6})_{c}\left(\xi_{1}\xi_{2}\right)}$ & 19814 & 653 & 534 & -785 & 4.39 & 0 & 0\tabularnewline
 & $1^{3}P_{2^{++}(6\bar{6})_{c}\left(\xi_{1}\xi_{3},\xi_{2}\xi_{3}\right)}$ & 19691 & 679 & 481 & -878 & -0.9 & -0.16 & 1.03\tabularnewline
 & $1^{3}P_{2^{++}(\bar{3}3)_{c}\left(\xi_{1}\xi_{3},\xi_{2}\xi_{3}\right)}$ & 19777 & 662 & 516 & -809 & -2.33 & 0.04 & 1.97\tabularnewline
 & $1^{3}P_{2^{++}(6\bar{6})_{c}\left(\xi_{1}\xi_{2}\right)}$ & 19891 & 644 & 562 & -721 & -5.38 & 0.21 & 3.13\tabularnewline
 & $1^{1}D_{2^{++}(6\bar{6})_{c}\left(\xi_{1},\xi_{2}\right)}$ & 19773 & 661 & 515 & -814 & 2.27 & 0 & 0\tabularnewline
 & $1^{1}D_{2^{++}(\bar{3}3)_{c}\left(\xi_{1},\xi_{2}\right)}$ & 19881 & 660 & 564 & -749 & -2.37 & 0 & 0\tabularnewline
 & $1^{1}D_{2^{++}(6\bar{6})_{c}\left(\xi_{3}\right)}$ & 19770 & 674 & 521 & -835 & 2.53 & 0 & 0\tabularnewline
 & $1^{1}D_{2^{++}(\bar{3}3)_{c}\left(\xi_{3}\right)}$ & 19806 & 666 & 530 & -796 & -2.9 & 0 & 0\tabularnewline
 & $1^{1}D_{2^{++}(\bar{3}3)_{c}\left(\xi_{1}\xi_{2}\right)}$ & 19697 & 695 & 484 & -880 & -9.67 & 0 & 0\tabularnewline
 & $1^{1}D_{2^{++}(6\bar{6})_{c}\left(\xi_{1}\xi_{2}\right)}$ & 19848 & 672 & 546 & -766 & -11.5 & 0 & 0\tabularnewline
 & $1^{3}D_{2^{++}(\bar{3}3)_{c}\left(\xi_{1},\xi_{2}\right)}$ & 19894 & 655 & 568 & -735 & -0.12 & 0.26 & -2.31\tabularnewline
 & $1^{3}D_{2^{++}(6\bar{6})_{c}\left(\xi_{1}\xi_{3},\xi_{2}\xi_{3}\right)}$ & 19910 & 654 & 576 & -727 & 0.96 & -0.62 & -2.24\tabularnewline
 & $1^{3}D_{2^{++}(\bar{3}3)_{c}\left(\xi_{1}\xi_{3},\xi_{2}\xi_{3}\right)}$ & 19862 & 657 & 552 & -751 & -1.56 & 0.47 & -2.46\tabularnewline
 & $1^{5}D_{2^{++}(\bar{3}3)_{c}\left(\xi_{1},\xi_{2}\right)}$ & 19881 & 660 & 564 & -749 & 3.45 & 0.32 & -5.86\tabularnewline
 & $1^{5}D_{2^{++}(\bar{3}3)_{c}\left(\xi_{3}\right)}$ & 19807 & 665 & 530 & -795 & 3.93 & 0.34 & -5.54\tabularnewline
 & $1^{5}D_{2^{++}(6\bar{6})_{c}\left(\xi_{1}\xi_{2}\right)}$ & 19856 & 659 & 550 & -759 & 3.36 & 0.42 & -6.83\tabularnewline
 &  &  &  &  &  &  &  & \tabularnewline
\multirow{7}{*}{$3^{+-}$} & $1^{5}P_{3^{+-}(6\bar{6})_{c}\left(\xi_{1}\xi_{2}\right)}$ & 19902 & 627 & 569 & -711 & 2.39 & 0.4 & 5.98\tabularnewline
 & $1^{3}D_{3^{+-}(\bar{3}3)_{c}\left(\xi_{1},\xi_{2}\right)}$ & 19887 & 651 & 568 & -744 & -0.42 & -0.01 & 3.81\tabularnewline
 & $1^{3}D_{3^{+-}(\bar{3}3)_{c}\left(\xi_{3}\right)}$ & 19812 & 657 & 533 & -789 & -0.61 & 0.12 & 3.62\tabularnewline
 & $1^{3}D_{3^{+-}(6\bar{6})_{c}\left(\xi_{1}\xi_{3},\xi_{2}\xi_{3}\right)}$ & 19917 & 643 & 581 & -720 & 0.64 & -0.01 & 4.36\tabularnewline
 & $1^{3}D_{3^{+-}(\bar{3}3)_{c}\left(\xi_{1}\xi_{3},\xi_{2}\xi_{3}\right)}$ & 19869 & 647 & 557 & -745 & -1.64 & -0.2 & 4.78\tabularnewline
 & $1^{3}D_{3^{+-}(6\bar{6})_{c}\left(\xi_{1}\xi_{2}\right)}$ & 19857 & 657 & 552 & -757 & -6.3 & -0.28 & 4.52\tabularnewline
 & $1^{5}D_{3^{+-}(\bar{3}3)_{c}\left(\xi_{1},\xi_{2}\right)}$ & 19901 & 645 & 573 & -729 & 3.03 & 1.05 & 0\tabularnewline
 &  &  &  &  &  &  &  & \tabularnewline
\multirow{6}{*}{$3^{++}$} & $1^{3}D_{3^{++}(\bar{3}3)_{c}\left(\xi_{1},\xi_{2}\right)}$ & 19901 & 644 & 573 & -729 & -0.12 & -0.07 & 4.5\tabularnewline
 & $1^{3}D_{3^{++}(6\bar{6})_{c}\left(\xi_{1}\xi_{3},\xi_{2}\xi_{3}\right)}$ & 19917 & 642 & 581 & -720 & 0.93 & 0.17 & 4.35\tabularnewline
 & $1^{3}D_{3^{++}(\bar{3}3)_{c}\left(\xi_{1}\xi_{3},\xi_{2}\xi_{3}\right)}$ & 19869 & 646 & 557 & -745 & -1.52 & -0.13 & 4.78\tabularnewline
 & $1^{5}D_{3^{++}(\bar{3}3)_{c}\left(\xi_{1},\xi_{2}\right)}$ & 19887 & 650 & 569 & -743 & 3.37 & 0.82 & 0\tabularnewline
 & $1^{5}D_{3^{++}(\bar{3}3)_{c}\left(\xi_{3}\right)}$ & 19813 & 655 & 534 & -788 & 3.83 & 0.88 & 0\tabularnewline
 & $1^{5}D_{3^{++}(6\bar{6})_{c}\left(\xi_{1}\xi_{2}\right)}$ & 19864 & 647 & 555 & -751 & 3.26 & 1.08 & 0\tabularnewline
 &  &  &  &  &  &  &  & \tabularnewline
\multirow{1}{*}{$4^{+-}$} & $1^{5}D_{4^{+-}(\bar{3}3)_{c}\left(\xi_{1},\xi_{2}\right)}$ & 19908 & 634 & 578 & -722 & 2.94 & -0.51 & 8.74\tabularnewline
 &  &  &  &  &  &  &  & \tabularnewline
\multirow{3}{*}{$4^{++}$} & $1^{5}D_{4^{++}(\bar{3}3)_{c}\left(\xi_{1},\xi_{2}\right)}$ & 19894 & 640 & 573 & -737 & 3.29 & -0.4 & 7.41\tabularnewline
 & $1^{5}D_{4^{++}(\bar{3}3)_{c}\left(\xi_{3}\right)}$ & 19819 & 645 & 538 & -782 & 3.75 & -0.43 & 7.02\tabularnewline
 & $1^{5}D_{4^{++}(6\bar{6})_{c}\left(\xi_{1}\xi_{2}\right)}$ & 19871 & 636 & 560 & -744 & 3.17 & -0.52 & 8.56\tabularnewline
\hline\hline
\end{tabular}
\end{center}
\end{table*}

%%%%%%%%%%%%%%%%%%%%%%%%%%%%%%%%%%%%%%%%

%%%%%%%%%%%%%%%%%%%%%%%%%%%%%%%%%%%%%%%%

%%%%%%%%%%%%%%%%%%%%%%%%%%%%%%%%%%%%%%%%

\section*{Acknowledgement}

This work is supported by the National Natural Science Foundation of China (Grants Nos. U1832173, 11775078, 12175065, 11425525, and 11521505). Q.Z. is also supported in part, by the DFG and NSFC funds to the Sino-German CRC 110 ``Symmetries and the Emergence of Structure in QCD'' (NSFC Grant No. 12070131001), the Strategic Priority Research Program of Chinese Academy of Sciences (Grant No. XDB34030302), and National Key Basic Research Program of China under Contract No. 2015CB856700.

% ²Î¿¼ÎÄÏ×%%%%%%%%%%%%%%%%%%%%%%%%%%%%%%%%%%%%%%%%%%%%%%%%%%%%%%%%%%%%%%%%%%%%%%%%%%%%%%%%%%%%%%%%%%%%%%%%%%%%%%%%%%%%%%%%%%%%%%%%%%%%%%%%%%%%%%%%%%%%%%

\bibliographystyle{unsrt}

\end{document}